# Paths to Scalable Carbon Neutrality for MIT:
Recommendations for MIT to stop contributing to climate change and inspire others at an affordable price

Caleb Amy, Julien Barber, and Colin Kelsall

May 16th, 2018

MIT 2.S999

# Introduction

The goal of this report is to provide actionable recommendations for paths to carbon neutrality from MIT's largest group of stakeholders, students. In the report, we consider the optimization of emissions reductions, the timescale of solution implementation, cost, scalability, and the associated risks inherent to two different scenarios. Similarly, it leverages MIT's strengths, such as its position as the world leader in technology and innovation as well as its financial strength, which enables it to make relatively low return investments and use contracting power to enable renewable projects.

The term "carbon neutral" has been used in many ways. We believe MIT will have achieved meaningful carbon neutrality when its net greenhouse gas (GHG) emissions reach zero as a result of employing scalable solutions with additionality. In our proposal, we consider a project 'additional' when it results in a tangible reduction of emissions that would not have occurred without the financial investment from MIT. Likewise, scalable solutions are critically important to directly powering our campus. For example, sources such as solar, wind, hydro and nuclear can be scaled to be key players in a future sustainable society, but current biofuels (which could be burned on campus) such as ethanol and wood pellets have too low of an energy production density to be scaled dramatically. In the same way, we assert that the rate of emissions reductions will be maximized if the lowest cost (per GHG emission avoided) approaches are taken first—which may not necessarily be in New England.

A critical aspect of renewable energy supply is its variability, or lack of dispatchability. This issue necessitates energy storage, demand response, and/or diverse supply. We argue that a combination of all three approaches is optimum, and that MIT should take actions to either produce or enable these approaches.

The timeframe and cost of recommendations provided in this report shift depending on the assumptions used. For this reason, key assumptions are explained here, and a more detailed explanation is given in Appendix A. First, we assume the Massachusetts Renewable Portfolio Standard (RPS) will continue indefinitely, at 1% per year increase of renewable generation. Moreover, we also assume that in 2017 the grid was 37% carbon free, based on 11% renewables, 6% hydro, and 20% nuclear. Although there is currently more nuclear generation in NE-ISO and its future is unclear, we assume grid nuclear capacity remains at a steady 20% in our models. This assumption results in a carbon free grid in 2080 (although we estimate the LCA emissions of this grid would be 7% of current emissions). We use a 5% internal rate of return for MIT. We evaluate proposed actions on cost per avoided carbon in Eq. 1, and detailed in Appendix A, Cost Modeling.

$$\frac{Cost}{Carbon} = \frac{CAPEX_{marginal} - \sum_{t=1}^{T} \frac{Cost\ Savings}{(1+r)^t}}{Lifetime\ Avoided\ Emissions} \quad (1)$$

This report begins by presenting a low risk Scenario A that focuses on implementing well understood technologies through a sustainability fund and PPAs to rapidly offset emissions in the near-term while gradually engaging in an ever-cleaner grid. Then, a more aggressive local scenario is proposed, where MIT leverages technological innovation in energy storage and HVAC systems to drastically reduce demand and load shift to accommodate variable renewables on the NE-ISO. These plans include an analysis of the cost, emissions, scalability, land use, and risk as well as the technological basis of the key technologies proposed. Finally the results of each plan are compared, and key recommendations are made.

## Scenario A: Rapid Carbon Neutrality PPAs and a Climate Action Fund

Climate change is a time dependent problem. Failure to act in the near term severely impacts our ability to mitigate the impacts of global warming in the future. As such, it is the moral, ethical, and social responsibility of global leaders across government, industry, and academia to step up and take action where action is possible. We believe MIT to be one such global leader with not only the economic resources to effectively invest in emissions reductions, but the world renown influence required to inspire action within communities around the world. For these reasons, we propose Scenario A as a strategic plan designed leverage MIT's influence to most effectively reduce global emissions on a time horizon representative of the urgent need for action.

Within Scenario A we propose MIT join the One Billion Dollar Green Challenge through the establishment of a revolving Climate Action Fund (CAF). Joining the One Billion Dollar Challenge would add MIT to the list of 62 trailblazing educational institutions across North America that are committed to taking financial action against climate change through the sustained investment structure of a revolving investment fund. From the CAF, initial funding will be provided for energy efficiency projects and other emissions reducing energy investments. Through a network of pre-investment measurement and data collection, the return from these investments are then returned to the fund to repay the initial capital and reinvest in future projects.

Applications to the investment fund are evaluated based on ROI, emissions impact, community engagement, ability to demonstrate verifiable savings, alignment with MIT's Climate Action Plan, and ability to enhance the effectiveness of education and research at MIT. Ultimately, the Climate Action Fund will be used as a mechanism to execute MIT's Climate Action Plan. Through the CAF, MIT can implement an effective shadow price for carbon within its infrastructure and efficiency investments. Moreover, the fund can be used as a tool to use the campus as a test bed for new technologies, and engage students in solving for carbon neutrality. We propose an annual student competition in which interdisciplinary teams of students will have the opportunity to pitch an investment project to the CAF board. An initial hackathon (or perhaps as a part of a class such as Solving for Carbon Neutrality [1]) can give students the opportunity to connect with relevant faculty and staff across the various departments to spark initial concept development. The top three ideas will then continue to develop their ideas with support from relevant institute partners. Each project should be subject to the same application conditions as any other CAF project. However, the ROI constraint should be relaxed in order to encourage student engagement and promote innovative ideas. The student competition allows MIT to continue to use the campus as a test bed for lab scale technologies, act as a runway for scalable sustainable solutions, and most importantly, engage the MIT community in working together to collectively find solutions to carbon neutrality.

Benchmarking against other universities involved in the Challenge, we propose MIT invest $9M annually in the fund, increasing the annual investment in infrastructure by $4M. The $9M annual investment is calculated based on a cost of carbon mitigation of $45/tCO2 [2] and MIT's annual CO2 emissions of roughly 200,000 tCO2. However, we have also benchmarked this investment against other Challenge participants which have a similar research profile as MIT. For reference, the California Institute of Technology committed $3M annually and saw a 22% reduction in emissions over 5 years. Caltech is roughly 1/6th the size of MIT in both population and annual energy use. Even with their lower investment and overall energy usage, Caltech saw a 26% ROI and $7M in avoided utility costs over their first 5 years. Through the CAF, MIT can establish the financial mechanism required to track, sustain, and accelerate short and long term investments in infrastructure and carbon-negative projects. The projects invested in by the CAF will ultimately reduce operating expenses through improved efficiency, create green jobs on campus, and provide sustained emissions reductions in service of our greater Climate Action Plan. In Scenario A, we propose investing in operational efficiency and building envelope improvements, on-site renewable energy, and off-site PPAs as potential opportunities for the CAF. As the carbon intensity of the grid improves, we then lay out plans for electrification of our major utility systems on campus.

### Efficiency Improvements

The first investment opportunities to consider are efficiency improvements that help MIT reduce its demand. Efficiency improvements are broad and should be evaluated on an individual basis through CAF. However, each project can be categorized as either infrastructure enhancements or digitization and automation of energy use. Through the One Billion Dollar Challenge, participating universities have published their successful energy conservation measures including but not limited to: installing motion sensors, high efficiency lighting, automated fume hood control, re-roofing buildings with reflective material, making LED signage standard throughout campus, increasing diameter of distribution piping, installing VFDs on all pumps, compressors, and fans, participating in local utility appliance replacement programs, and improving air leakage from windows and building envelope [3].

Caltech, a similar energy dense, urban research institute reported an 18% emissions reductions through their efficiency measures due to their revolving fund over the first five years. Considering the age of the MIT campus infrastructure and the scale of energy use compared to Caltech, we conservatively estimate our emissions reductions at 12% of the 2014 baseline.

### On-Site Renewable Energy

In addition to efficiency improvements, certain on-site renewable projects can be implemented through the CAF investments. Rooftop solar currently present the most feasible option for MIT's dense urban campus. Utilizing ~20% of the available rooftop space on campus, and with a solar potential of 35 W/m$^2$ in Cambridge, MA, we would be able to reduce our emissions by ~5% of the 2014 baseline [4]. The total system today would cost approximately $4.6M [5], with an NPV cost of abated carbon of $15/tCO2. As the price of solar drops, local generation on MIT rooftops will become increasingly cost competitive and reduce the cost of abated carbon in subsequent lifetime investments. Moreover, making on-site renewable generation visible and accessible to members of the MIT community has the added benefit of increasing awareness of the Institute's Climate Action Plan among the faculty, staff, and students.

### Off-Site Power Purchase Agreements

Further, we propose that renewable power purchase agreements (PPAs) be an integral part of the CAF. Granted, investments in PPAs may not contribute to direct carbon emissions reductions on campus. However, these financial contracts allow MIT to tangibly reduce global emissions in the near-term through the execution of additional renewable energy projects in locations where carbon-free energy makes the biggest difference. Climate change is not a local problem. If our goal is to contribute to the fight against global warming, then it is our moral obligation to leverage the least-cost solutions that most effectively reduce global emissions while still serving the greater interests of the Institute. Moreover, we must consider time as a factor. Climate change is happening now and we need solutions that reduce substantial emissions now. As a recognized global leader, MIT has the resources to step up and generate solutions to our climate problem where others may not.

Integrating PPAs into the CAF provides a dedicated investment fund to substantially invest in near-term, carbon-reducing projects that provide the least-cost solution to reducing overall emissions. As an integral part of the CAF, the net returns will be added to (or taken from) the CAF. By investing in PPAs now, we will be able to capitalize on the positive financial returns in the next 10-20 years in order to contribute to the available capital to invest in efficiency and infrastructure projects here on campus. However, we recognize that there are risks associated with PPAs, and therefore account for their financial impact as net neutral – i.e. PPAs don't make or cost us any money.

Through the solicitation of PPAs, proposals are to be carefully screened to mitigate the risks to the Institute. The two main risks to consider relate to the economic and reputational risk exposure inherent to each project. From an economic perspective, PPAs should be selected that have a high probability of providing positive financial return to the CAF. Understanding market and regulatory factors, scheduled generation facility retirements, new generation construction, facility load profile, transmission planning, and past and future policy decisions can help to ensure a profitable contract. Moreover, ensuring that developers have land, permits, and interconnection agreements are positive signs that the project will meet its projected targets.

From a reputational perspective, ensuring the selected projects are certified additional is essential to meeting MIT's Climate Action goals. The purchase of non-additional RECs is at best a small subsidy for renewable projects. To seriously engage in emissions reductions, and to serve as a trusted leader in corporate and institutional sustainability, we must be sure to only claim credit where credit is due. Using third party additionality verification can be an effective means to ensure our investments are having the greatest impact possible. Moreover, the projects we invest in are a reflection of the Institute itself. The developers we purchase power from represent MIT during the construction and operation of the asset. Therefore, it is important to screen developers based on their past project development record. Only developers with a strong completion record, as well as positive reputation amongst their suppliers and the communities in which they operate, should be selected to conduct business with MIT.

MIT has the opportunity to use its resources and convening power to drive the clean energy transition in a time when it is desperately needed. We believe that it's the responsibility of the Institute to invest in measures that result in the largest amount of emissions reductions per dollar spent. The use of PPAs in scenario A allow us to leverage these near-term emissions investments and take advantage of the current favorable economics associated with renewable PPAs to fund our local efficiency improvements. Although we propose the use of off-site PPAs, we recognize that local and regional measures will be critical to becoming truly carbon neutral, and this is part of our approach as well.

## Electrification of Campus HVAC

When the new central utility plant (CUP) system approaches end of life in 2040, MIT will be faced with a decision whether to continue operating an on-site combined heat-and-power plant or to rely on the grid for the full energy load. We predict the 2040 grid carbon intensity to be 200 g$CO_2$/kWh, lower than the new CUP's rated heating carbon intensity of 227 g$CO_2$/kWh [6]. To take full advantage of the decarbonizing ISO-NE grid, MIT needs to completely electrify the campus by the retirement date of the new CUP. Currently, the campus uses natural gas combustion in some part for all three utility streams: steam, electricity, and chilled water indirectly with steam-powered chillers. In this scenario, we propose well established technologies to integrate into the grid as described in the following sections. Table 1 summarizes the relevant economic and carbon mitigation metrics for the full electrification of campus HVAC.

**Table 1:** Summary of Scenario A campus HVAC electrification evaluation metrics

| | |
|---|---|
| Proposed Implementation Date | 2040 |
| Total Capital Cost | $335 million |
| Annual cost savings ($/year) | $10 million |
| Annual mitigated emissions (t$CO_2$/year) | 76,000 |
| NPV cost per annual mitigated emissions ($/t$CO_2$-year) | $2,000 |
| NPV cost per total mitigated emissions ($/t$CO_2$) | $50 |

### Steam to hot-water conversion

An important enabling technology for the electrification of the current heating system is the conversion from steam to hot-water. Producing steam from electricity typically requires the implementation of electric boilers with a maximum $COP_h$ of 1. By switching from steam lines at 200°C/400°F [6] to a hot-water distribution network at 75°C/170°F, heating can be provided using heat pumps with coefficients of performance (COP) of 2 or greater [7]. COP is defined as the ratio of heating power to input electrical power, as a COP of 2 is effectively an electricity-to-heat conversion efficiency of 200%. Besides enabling this higher efficiency technology, the lower temperature water circulation and simplified equipment typically result in more efficient distribution. In 2015, the University of British Columbia completed an ambitious steam to hot-water conversion project that is predicted to decrease the heating load by 26% through the reduction of distribution system and end user losses [8]. Figure 1 shows where these improvements come from and how they can drastically improve the overall heat distribution efficiency. It is important to note, that although UBC switched from steam to hot-water through this project, they maintained their existing system of boilers for water heating and did not take advantage of the additional benefits of switching to heat pumps at this time.

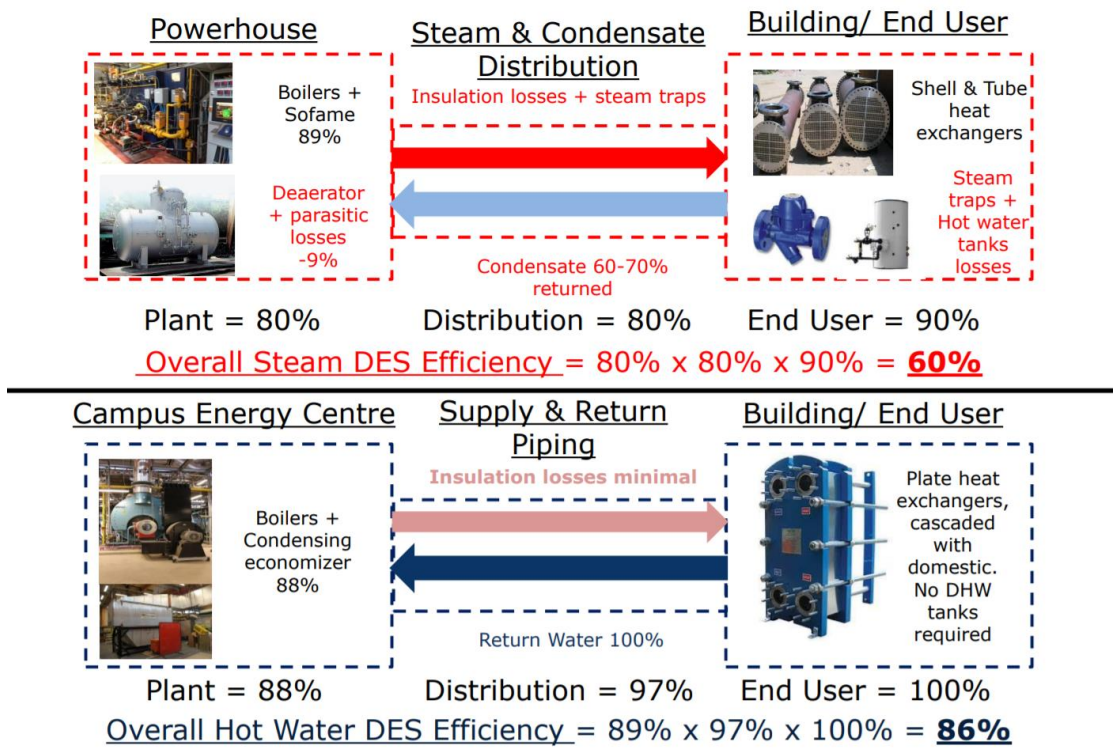

**Figure 1:** System efficiency improvements predicted from steam to hot-water conversion at the University of British Columbia [8]

We propose the conversion of campus from steam to hot water by 2040, when the new CUP is retired and we no longer have a significant waste heat stream available. If a system similar to UBC's were implemented at MIT, it could reduce our heating load by 26% and our emissions by approximately 15,000 $tCO_2$/year based on MIT's predicted heating load in 2040. This calculation assumes that the carbon intensity of the heating provided by MIT's new CUP is 207 $gCO_2$/kWh$_t$ [6], the campus heating load increases by 10% every 20 years, and that a 12% load decrease results from efficiency improvements over the next 10 years. Mitigated emissions were calculated using the CUP intensity and predicted heating efficiency improvements, this assumes that no additional electrical heating facilities are implemented. The total lifetime net emissions abatement contribution of the steam to hot water conversion over a 40-year lifespan was calculated to be 625,000 $tCO_2$.

MIT estimates the cost of implementing a similar conversion on campus at approximately $225 million [9], which results in a cost per carbon abated per year of $12,000/tCO$_2$-year and a cost per total carbon abated of $280/tCO$_2$. Although these costs are relatively high when compared to our other recommended carbon mitigation strategies, this conversion is a critical *enabling* step towards electrifying campus and realizing the reductions available on that path.

The following additional assumptions were used to calculate the cost per mitigated emissions: discount rate of 5%, gas heating efficiency of 90%, and natural gas cost of $10/MMBTU. Table 2 summarizes mitigated emissions, costs and savings.

**Table 2:** Summary of steam to hot-water conversion evaluation metrics

| Proposed Implementation Date | 2040 |
|---|---|
| Capital Cost | $225 million |
| Annual cost savings ($/year) | $2.8 million |
| Annual mitigated emissions (tCO$_2$/year) | 15,000 |
| NPV cost per annual mitigated emissions ($/tCO$_2$-year) | $12,000 |
| NPV cost per total mitigated emissions ($/tCO$_2$) | $280 |

### Electric heating

Once MIT has transitioned to a hot-water based heat distribution network and the combined heat and power CUP has been retired, electrically powered heat pumps can be implemented to take advantage the inherent performance benefits. Although ground-source heat pumps (GSHP) typically obtain much higher COPs than their air-source (ASHP) counterparts [10], for this conservative scenario we considered the added cost and disruption of GSHPs to be too great and have only evaluated the potential for air-source options. Still, a primary concern with the implementation of ASHPs in cold climates is that their performance is highly dependent on the exterior air temperature. Because of this, we propose an ASHP that uses $CO_2$ as the working fluid instead of a typical refrigerant. Temperature dependent performance of an example $CO_2$ based ASHP water heater produced by Sanden USA and capable of heating water to 170°F/75°C is shown in Figure 2A.

The average low daily temperatures in Boston from 1981-2010 [11] are shown in B, along with the predicted Sanden SANCO2 ASHP COP using the manufacturer's specifications. Based on this analysis, there are no average days where the heat pump COP would drop below 2. For those few very cold days where the heat pumps are not able to keep up with demand, some electrical resistive heating may need to be used in addition to the pumps.

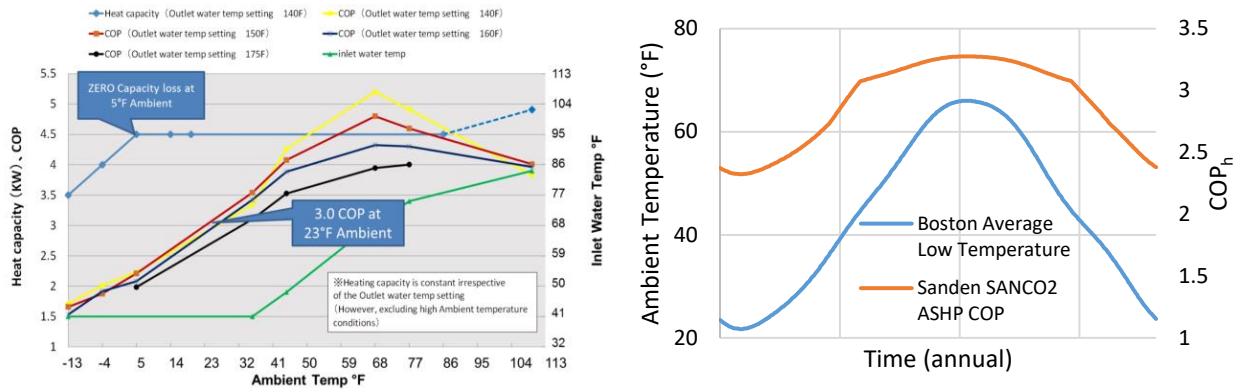

**Figure 2: A** – Performance data for a $CO_2$ air-source water heater capable of producing 170°F/75°C water at outside air temperatures as low as -13°F/-25°C [7]. **B** – Boston average daily low temperature 1981-2010 [11] and predicted $CO_2$ ASHP COP

**Table 3:** Summary of electric heating (ASHP) evaluation metrics

| Proposed Implementation Date | 2040 |
|---|---|
| Capital Cost | $110 million |
| Annual cost savings ($/year) | $5.3 million |
| Annual mitigated emissions (t$CO_2$/year) | 47,000 |
| NPV cost per annual mitigated emissions ($/t$CO_2$-year) | $260 |
| NPV cost per total mitigated emissions ($/t$CO_2$) | $6 |

Calculating the costs of such a system is challenging as $CO_2$ ASHPs are a relatively new technology, but the Sanden SANCO2 ASHP water heater cost approximately $900/kW$_t$ [12]. Based on MIT's average monthly heating load in 2017 [13], an average heating power requirement can be obtained for each month. The maximum load was in February 2017 and resulted in an average heating load of 60 MW$_t$. To account for peaks in demand, a factor of 2 was used when sizing an ASHP system. To provide 120 MW$_t$ of heating would require a system that costs roughly $110 million, as shown in Table 3. Additional construction and installation costs were unknown and neglected for this preliminary model.

### Electric cooling

Switching to a fully electric cooling system is a much simpler prospect. Electric chillers are already implemented in the current CUP system and are utilized when the chilled water load exceeds the capacity of the steam-powered chiller. Approximately 40% of the annual chilled water load is provided by electric chillers (COP$_c$=5.4) while the other 60% is provided by steam-powered chillers (COP$_c$=1.8). If MIT switches to fully electric chillers by 2040, taking full advantage of the high COP$_c$ of the electric units as well as the decarbonizing grid, roughly 13,000 t$CO_2$ of emissions per year could be mitigated with a cost savings of $2 million per year. Assuming a 40-year lifespan of the new chiller units and accounting for grid decarbonization, the switch would result in total lifetime mitigated carbon emissions of 550,000 t$CO_2$.

When considering the cost of switching to a fully electric system, the total capacity of the current steam chillers was replaced with electric chillers. We assumed that industrial scale water chillers cost approximately $300 per ton of capacity [14], and MIT's current steam chillers have a

10,500-ton capacity. Assuming complete replacement, we calculate the conversion to cost around $5 million for the chillers alone. Based on Eq. 1, cost per avoided carbon is estimated to be $(55)/tCO$_2$. A negative number indicates MIT would make money from this action. The results of this financial analysis are presented in Table 4.

**Table 4:** Summary of electric chiller conversion evaluation metrics

| Proposed Implementation Date | 2040 |
|---|---|
| Capital Cost | $5 million |
| Annual cost savings ($/year) | $2 million |
| Annual mitigated emissions (tCO$_2$/year) | 13,000 |
| NPV cost per annual mitigated emissions ($/tCO$_2$-year) | $(2,800) |
| NPV cost per total mitigated emissions ($/tCO$_2$) | $(55) |

The overall process of achieving carbon neutrality in Scenario A is presented in Figure 3. As shown, carbon neutrality is achieved by 2020, and is maintained indefinitely by ever smaller PPAs, as demand is reduced and the grid becomes cleaner. While Scenario A is effective at achieving the key goal of a carbon neutral MIT, it doesn't fully leverage one of MIT's greatest strengths: innovation. For this reason, we present Scenario B that exchanges the leveraging of MIT financial strength with leveraging its power to innovate and drive technology adoption.

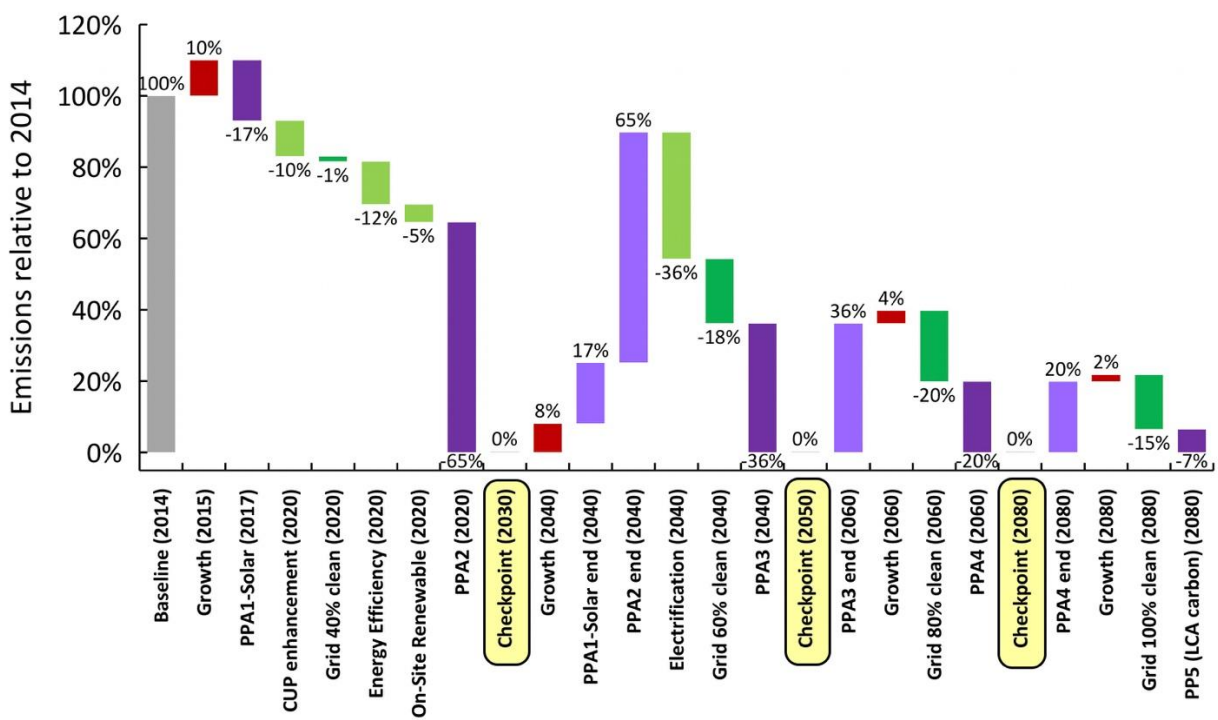

**Figure 3:** Summary of Scenario A proposed MIT emissions elimination strategy

## Scenario B: Advanced Efficiency and Energy Storage

In Scenario B we rely on technology and innovation to reach carbon neutrality without off-site offsets via PPAs. Although this approach takes longer, and is more expensive, we believe it is worth considering since it leverages MIT's strength of spearheading technological innovation and inspiring others. Here, we propose more advanced HVAC improvements and the use of energy storage to shift our demand to reduce cost and emissions, while helping to enable variable renewables.

## Electrification of Campus HVAC

Our more ambitious approach to the electrification of campus HVAC system follows a similar process as Scenario A. That is, it includes conversion of steam to hot-water and the electrification of the chillers and heaters. Like many large institutions, MIT currently produces steam and chilled water simultaneously but with independent systems. In this section, we analyze the potential of a heat-recovery chiller coupled to the heating system as an avenue for efficiency improvements. Additionally, we propose the addition of hot and cold-water storage which would allow MIT to shift the electrical load to optimal times during the day, taking advantage of period with cleaner power and lower prices. An overall summary of the results of Scenario B are provided below in Table 5.

**Table 5:** Summary of Scenario B campus HVAC electrification evaluation metrics

| Proposed Implementation Date | 2040 |
|---|---|
| Capital Cost | $560 million |
| Annual cost savings ($/year) | $6.5 million |
| Annual mitigated emissions (t$CO_2$/year) | 46,000 |
| NPV cost per annual mitigated emissions ($/t$CO_2$-year) | $9,600 |
| NPV cost per total mitigated emissions ($/t$CO_2$) | $240 |

### Heat-recovery Chillers with thermal storage

After the retirement of the CUP in 2040, MIT will no longer be able to use turbine waste heat for all our heating needs. As such, capturing any streams of waste heat will become increasingly important. Heat recovery chillers (HRCs) are a useful option for capturing the heat generated by chilling water that is normally expelled through a cooling tower. Figure 4 indicates a potential configuration of a heat-recovery chiller and thermal storage tanks to account for hourly fluctuations of relative heating and cooling demand.

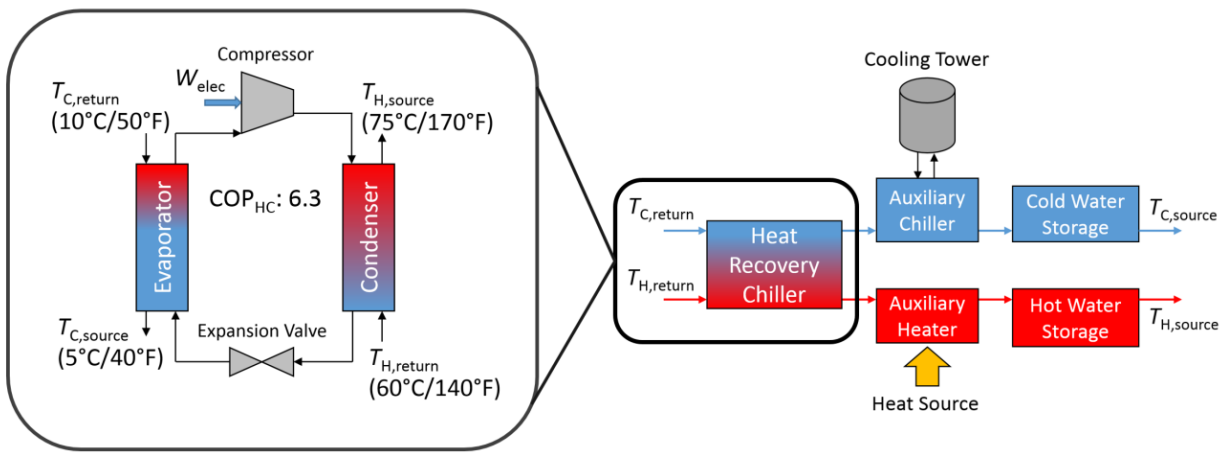

**Figure 4:** Potential configuration of a HRC within a heating and cooling system

In 2015, Stanford University implemented the Stanford Energy System Innovations (SESI) facility that takes advantage of simultaneous heating and cooling on an institutional level. The SESI system captures 53% of the waste heat from cooling and uses it to provide 88% of their annual heating load. They also integrate 315 MWh of cold water storage and 175 MWh of hot water storage for load leveling and shifting. The project cost $438 million in total and included the conversion of campus from steam to hot-water. Through the implementation of this combined system, the university was able to reduce their carbon emissions by 68% when compared to the gas-fired cogeneration previously in use [15].

To determine if such a system would make sense at MIT, hourly heating and cooling demand data for the entire year of 2014 [16] were analyzed with the results shown in Figure 5. In 2014, 88% of MIT's cooling waste heat could have been utilized to provide 46% of the annual heating load.

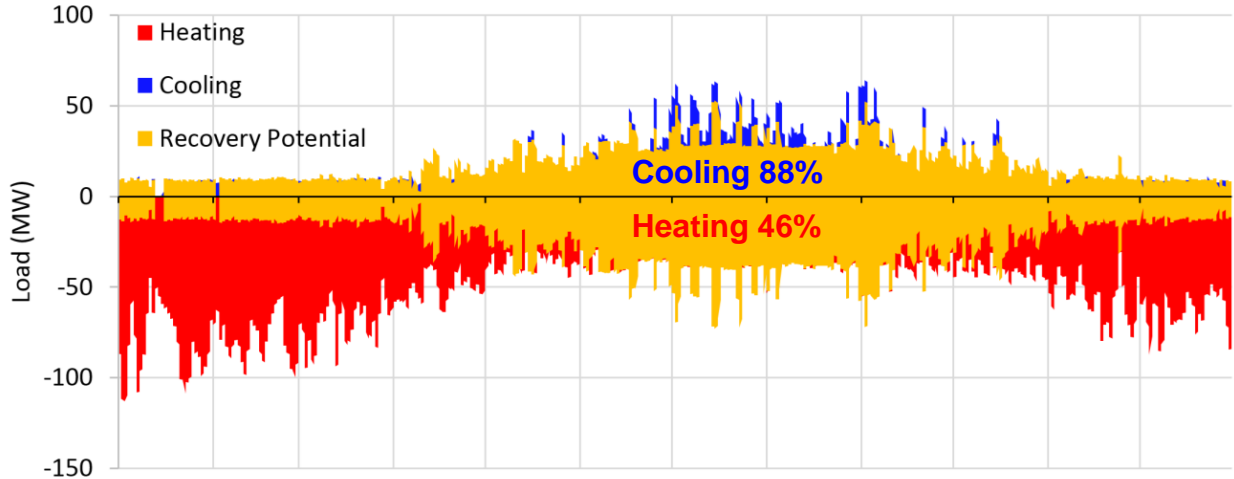

**Figure 5:** Evaluation of heat-recovery potential for MIT's hourly heating and cooling loads in 2014

To calculate the potential emissions reductions due to the implementation of a HRC at MIT in 2040, the remaining heating and cooling loads were assumed to be addressed by the pre-existing electric chillers and the ASHPs described in Scenario A. Heating and cooling demand was assumed to grow by 10% every 20 years, and a lifetime of 40 years was assumed for all equipment. Additionally, the 26% reduction in heating load due to a steam to hot-water conversion was incorporated into the calculation. The baseline case for comparison was the current hybrid electric ($COP_c$=5.4) and steam chiller ($COP_c$=1.8) systems, as well as the natural gas boilers for steam generation that have a 90% efficiency. Based on these assumptions, an HRC implemented in 2040 could mitigate 46,000 $tCO_2$/year of emissions and 1.8 million $tCO_2$ of the lifetime of the system.

When calculating the capital cost and savings for this proposed facility, the Stanford system was used as a baseline case. The SESI system cost $438 million in total, but the new HRC

power plant facility only accounted for 50% of the cost, $219 million [17]. In Stanford's system, the waste heat from cooling accounted for most of their heating load so there was no need to purchase significant capacity in additional electric heating systems. In MIT's case, however, the heating load still remains significant and additional ASHP system will be required to make up the difference. To account for this added cost in MIT's system, the capital cost of the proposed ASHP system in Scenario A ($110 million) was included. Finally, with the cost of steam to hot-water conversion included at $225 million, the total capital cost for the implementation of a fully electric hybrid HRC-ASHP heating and cooling system is roughly $560 million. The efficiency improvements realized with this conversion also result in financial savings of around $6.5 million per year. The NPV cost per mitigated carbon per year is roughly $9,600/t$CO_2$-year and the NPV cost per lifetime mitigated carbon comes out to $240/t$CO_2$. The results of this financial and emissions analysis are presented in Table 5.

When comparing this system to the facilities suggested in Scenario A, the inherent limitations of an HRC become obvious. The combined $COP_{hc}$ of the HRC installed by Stanford University was 6.3, which results in an effective $COP_h$ of 3.65 and $COP_c$ of 2.65. The $COP_c$ of MIT's current electric chillers is 5.4 and the $COP_h$ of the proposed ASHPs are around 2, these values result in HRCs only really being beneficial for heating applications. As seen in Table 5, using an HRC at full capacity (88% of cooling, 46% of heating) is not economical when compared to the independent heat pumps and electric chillers presented in Scenario A. There may be a portion of the year, however, during very cold weather where utilizing an HRC could significantly boost the performance of the campus heating system. During certain days were ASHPs would perform with a $COP_h$ of less than 2, utilizing an HRC for a portion of the heating load could be economical. Additionally, MIT should continue to evaluate technological developments as we electrify the campus to minimize the electricity required to condition the campus spaces.

### Heat Recovery

The majority of MIT's energy consumption is for heating and cooling, and this conditioned air is rapidly exhausted from campus buildings, especially in wet labs which require more than six exchanges of the lab's air volume per hour. It is primarily for this reason that wet labs consume 5-10x more energy than office space [18]. There is huge potential for energy recovery here, which has been recognized [19] and regulated for new buildings: 50% heat recovery is required on new projects, and MIT exceeds this. However, there is still major potential for retrofitting existing lab exhaust with heat recovery. We estimate, based on existing systems, that 2/3 of lab exhaust heat can be recovered, resulting in a 17% reduction in MIT's heating and cooling demand [18, 20, 21]. The cost of retrofitting existing buildings is estimated based on similar projects at other universities [22].

### Why Store Energy

As variable renewables come on to the ISO-NE grid, supply will become increasingly less able to respond to demand. When supply from renewables drops, the difference can be made up by peaking natural gas plants, but over time this will become costly as these expensive assets are only used for a few hours a day, and eventually only during a few hours a year during period of low renewable output. However, as it becomes difficult for supply to respond to demand, an alternative solution is mediation via energy storage. In this approach, thermal or electrical energy can be stored when renewable production is above demand and consumed when supply is low.

In MIT's case, storing energy enables several important benefits—which our neighbor Harvard has already acknowledged [23]. As wind, solar, hydro, and nuclear power all have very low variable cost, they will always be deployed when available, unlike fossil sources that have to buy fuel. There are two important consequences of this fact. First, if MIT decouples its electricity demand from supply, it can purchase this zero-carbon energy when it is available. Second, this energy will always be the lowest cost energy because energy prices reflect marginal, or variable costs, which are low for these sources. In fact, after approximately 2050, without energy storage or other demand response, renewable energy will begin to be curtailed as shown in Figure 6A. Future energy markets are expected to evolve [24] as a result of non-dispatchable renewable supply, but will continue to value energy storage.

Another benefit of energy storage is that it reduces MIT's exposure to electricity price spikes, which occur in highly variable grids. Similarly, and even more importantly, on campus energy storage provides resilience in the case of a grid failure. This enables MIT to retain some of the resilience the Campus Utility Plant (CUP) currently provides.

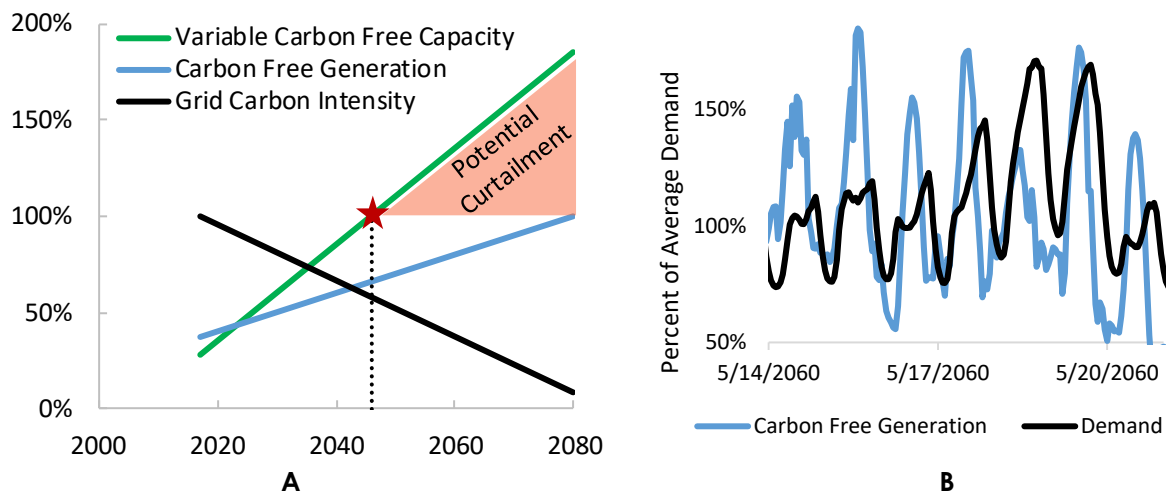

**Figure 6:** Future Massachusetts Grid under current RPS. A: After 2050, Renewable capacity exceeds average demand, so significant curtailment occurs unless storage or demand response is implemented. B: Typical week in 2060; when carbon free generation exceeds demand, energy is stored, consumed by flexible demand, or curtailed.

## MIT as an Electricity Storage Test Bed

The path to an electricity storage technology with costs low enough to enable deep decarbonization of the grid via renewables is unclear [25]. While many technologies exist, and even more are under-development, none meet the requirements that the future grid will have. For example, although pumped hydropower (PH) and compressed air energy storage (CAES) are both low cost technologies [26-28], especially on a cost per energy (CPE) basis, they are dependent on specific geographic locations, and therefore at best will only provide *some* of the storage needed. Geographically independent storage technologies, such as lithium ion and lead-acid batteries, flywheels, and others have fundamental costs that, when coupled with future solar and wind costs, will not fall below current fossil generation [29, 30].

There exist, however, exciting new energy storage ideas that could meet these requirements—and many of these are already under development at MIT. Some of the most promising areas of low cost energy storage involve storing electricity as heat or liquid electrolyte in flow batteries [29, 31, 32]. Another interesting concept, first proposed by Dr. Alex Slocum (MIT) [33], combines the approach of pumped hydro with the more available pressure difference in large bodies of water—mainly oceans. In this concept, a concrete sphere of 30m diameter located 700m below sea level stores energy by pumping water out of the sphere when excess power is available, and allowing it to fill later to generate power, as detailed in Appendix B1. The cost is expected to be significantly lower than the future cost of batteries [30, 34].

Given the benefits of on campus energy storage for MIT, the need to further develop this technology, and MIT's aptness to solve challenges of this type, we propose the use of MIT as a test bed for energy storage. Here, we propose to go beyond our current research stage, and actually *use* these new technologies to store energy for the campus. Although we propose starting this process as soon as possible, we see the ~2040 end of life of the CUP Titan 250 turbines as an opportunity to repurpose this space for energy storage. In fact, the technologies mentioned above have energy densities above 300 kWh/m$^3$, so 8 hours of MIT energy demand can be stored in less than 300 m$^2$, which is about 10% of the land area the CUP occupies. Further, this is not a new idea; in fact, the CUP has already engaged with students to demonstrate a thermal storage technology. We propose to grow this collaboration by adding space, electrical interface infrastructure, and competitive funding for labs and startups to demonstrate storage technologies.

The steps towards carbon neutrality proposed in Scenario B are summarized in Figure 7, along with their relative impacts on MIT's carbon emissions. As shown, carbon neutrality is achieved by 2050 without any PPAs and through entirely local initiatives. This scenario leverages MIT's technical strengths and focuses on developing new technologies to address carbon neutrality on a larger scale. While not all the elements proposed in this Scenario are currently economical or feasible, we tried to identify some areas where MIT could demonstrate its leadership in innovation and make it easier for others to follow suit.

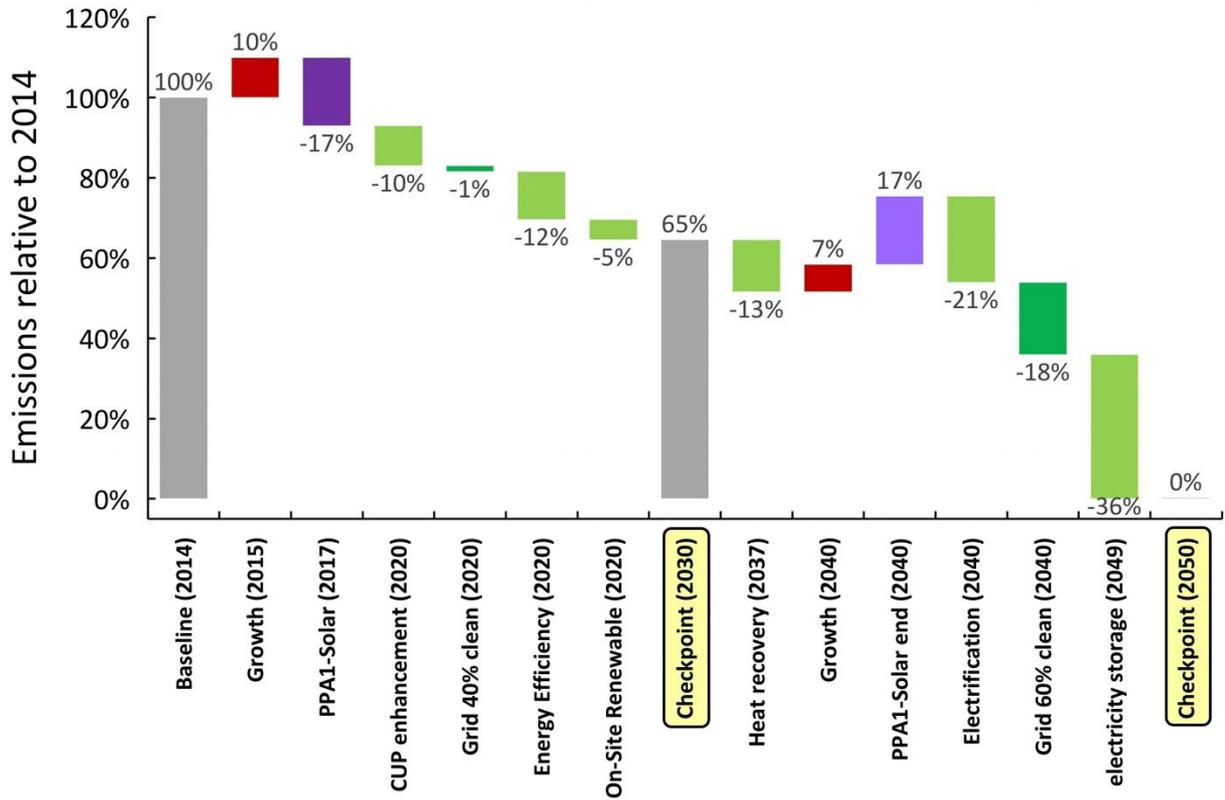

**Figure 7:** Summary of Scenario B proposed MIT emissions elimination strategy

## Conclusion

In this report we have presented two distinct approaches to bring MIT to carbon neutrality, but our ideal scenario would combine them. We believe MIT has the moral obligation to reach carbon neutrality as quickly as possible, and that carbon-free PPAs are the best way to achieve this. However, MIT should also demonstrate key technologies to reduce and shift demand, such as drastic efficiency improvements (e.g. simultaneous heat pump and heat recovery) and energy storage.

Although it may seem the best way for MIT to be carbon neutral is to produce its own energy from carbon-free sources, this approach is ineffective from cost and emissions perspectives. As explored in the report in Appendix B1, if MIT were to build a carbon-free microgrid, the lack of diversity would result in so much overbuilding of generation capacity and storage that energy costs would increase by 10X, with modest emissions reductions due to LCA emissions of the poorly used infrastructure.

On the other hand, in a diverse, fully utilized macrogrid, LCA emissions are expected to be 7% of current ISO-NE carbon intensity. To eliminate these emissions and reach true carbon neutrality would require indefinite offsets, or carbon sequestration. Appendix B1 includes an analysis of the LCA emissions of wind, solar, and several energy storage technologies.

Notably, scope 3 (indirect non energy) emissions are not addressed. Some approaches to reduce scope 3 emissions include further subsidizing sustainable transport, banning/taxing the sale of beef and the meat of other ruminates because of the related methane emissions, and using MIT's technical strength to solve the challenge of carbon-free global transport. These approaches were not proposed here because they are difficult to model, and cannot lead to carbon neutrality, only partial reduction in emissions.

A combination of these two plans may be the optimum path forward. For example, combining the solutions proposed in Scenarios A and B allow us to minimize cost by reducing demand up front, thereby reducing future CAPEX on energy projects and collecting energy savings indefinitely. Similarly, there is less social/political risk to engaging in PPAs if we are also taking drastic measures to reduce our local footprint. Lastly, uncertainty is reduced by spreading our approach over several aspects. For example, smaller PPAs can be coupled with energy storage to reduce exposure to future electricity price changes. Ultimately, the combination of these proposed solutions will serve to reduce both our local and global emissions, and help us lead the way in Institutional climate action.

# Appendix A

Cost Summary

| Item | Cost | Life (yr) | Lifetime Avoided Carbon | Discounted Savings (5% IRR) | Cost per Carbon ($/MTCO2e) |
|---|---|---|---|---|---|
| PPAs | $0 | 20 | 2000000 | $0 | $0 |
| On-Site PV | $4,600,000 | 20 | 45990 | $3,900,000 | $15 |
| Steam-->Hot water | $225,000,000 | 40 | 600000 | $57,000,000 | $280 |
| Electric Chillers | $5,000,000 | 40 | 520000 | $33,400,000 | -$55 |
| Air Source Heat Pump | $110,000,000 | 40 | 1880000 | $98,700,000 | $6 |
| Simultaneous Heat Pump | $335,000,000 | 40 | 1600000 | $75,900,000 | $162 |
| Heat recovery | $15,000,000 | 20 | 318000 | $27,000,000 | -$38 |
| Electricity storage (Future Li-Ion) | $26,400,000 | 10 | 385200 | $11,000,000 | $40 |

Assumptions

- Grid adds 1% renewable energy per year per Massachusetts RPS, starting at 37% in 2014, based on 20% nuclear and 6% hydro, which remain fixed.
- Cost per carbon is based on the lifetime avoided emissions of an action. This method is used to that the cost can be compared to MIT's calculated social cost of carbon.
- 5% discount rate for MIT investments used to discount future savings
- Natural Gas price of $10/MMBTU
- Electricity cost of $40/MWh$_e$

## Cost Modeling

The time value of money (TVM) is critically important to this analysis, especially when comparing solutions with different life. Cost figures reported here are based on the net present value (NPV) shown in Eq. 1. Here, future savings are discounted to today using a 5% internal rate of return (IRR). These savings are compared against to capital expenditure (CAPEX) to implement the item. Notably, in some case only the marginal cost is considered. For example, if a chiller must be replaced, then only the additional cost of installing a more efficient chiller is compared against the energy savings of that action. All NPV's are nominalized by emissions reduction, so that suggestions can be compared.

For example, because batteries last approximately half as long as StEnSea, they encounter a lower financing cost. This, fundamentally, is because instead of purchasing a 20-year plant on day 1, a 10-year plant is purchased. Then, 10 years later, the second half of capital expense (CAPEX) is required. For an interest rate, or IRR, of 7% this gives batteries a 25% benefit compared to purchasing two sets of batteries up front. Similarly, the fact that the StEnSea turbines are replaced after 20 years, compared to installing 2 sets up front, saves 38%—which is greater than the effect on batteries because of the longer time frame.

# Appendix B1, 2, 3

Previous Team Member Reports:
B1: Renewable Microgrid, Caleb Amy
B2: Power Purchase Agreements, Julien Barber
B3: Thermal Energy Storage, Colin Kelsall

# A Path to Scalable Carbon Neutrality for MIT

Caleb Amy

April 2nd, 2018

MIT 2.S999

# Introduction

This report begins by summarizing the status quo at MIT, and its current emissions reduction plan through 2030. Then, demand reductions and flexibility are proposed through 2045. Next, a fully off-grid low carbon scenario is proposed for MIT, including an analysis of the cost and emissions impact of the change. This plan is supplemented by additional approaches, of increasing diversity of supply. Following these plans, the basis of the key technologies proposed is then discussed and the methods used to compute costs are presented. Finally results of each plan, including a comparison to mainstream technologies, is included with a conclusion.

A critical aspect of renewable energy supply is its variability, or non-dispatchability. This issue necessitates energy storage, demand response, and/or diverse supply. In this report, an MIT microgrid supplied only by wind is compared to one supplied by wind and solar. Finally, the benefits of being connected to a diverse, clean macrogrid is discussed, and the benefits are projected.

The term "carbon neutral" is often defined liberally to include financial instruments and other methods, some of which are not scalable. Here, carbon neutral approaches are limited to those that that can be sustainably scaled to provide at least 20% of global energy demand. This limit is chosen because the final solution will use multiple sources, but mitigation approaches that cannot be scaled widely are not useful as a demonstration for others to follow. For example, sources such as solar, wind, and nuclear can be scaled to be key players in a future sustainable society, but biofuels such as ethanol and wood pellets have too low of an energy production density to be scaled dramatically. Similarly, approaches that "purchase" existing sustainable generation by financial instruments alone do not qualify. However, in instances where the emitter finances the construction of new and scalable sustainable generation, the offset may, in fact, influence the carbon intensity of human activity—with the greatest effect when fossil sources are directly replaced. To refer to approaches that include offsets, I use the term: carbon neutral with offsets. These have value, but not on a 1-1 level with sustainable production or demand reduction.

# The Current Plan at the Massachusetts Institute of Technology

MIT devotes a small minority (only one page) of its Climate Action Plan to emission reductions. Now, with MIT's position as *the* world leader in engineering and innovation, it is reasonable for it to focus on what it can contribute intellectually to the challenge of global warming. At the same time, however, global warming is as much a sociopolitical problem as a technical problem; it, therefore, requires both technical and sociopolitical approaches. A major goal of the ACUPCC, of which MIT is not a participant, is to be a "high-visibility" example of sustainable practices.

MIT sets the key goal of 32% emissions reduction by 2030 from 2014 levels[1]. The Institute plans to meet these emissions reductions despite anticipated space and program growth of 0.6% per year. To reach this goal, MIT has nearly eliminated relatively high emission fuel oil from use in its on-site 21 MWe CHP plant. MIT is also in the process of replacing its CHP plant with a new 44 MWe CHP plant, which is forecasted to reduce emissions by 10%, which would cancel the 10% increase expected from MIT Nano and other current growth projects, which will be completed around the same time (2019). This reduction is possible because the new plant will be more efficient and provide more heat for buildings, so that fuel oil and natural gas will be used less in boilers for heat only generation. An unfortunate consequence of MIT's decision to invest in new fossil power is that it essentially guarantees a floor of emissions through 2040, the life of the turbines.

On the demand side, MIT plans to eliminate 15% of 2014 emissions through energy efficiency investments in existing buildings. In its recent report "From Plan to Action", MIT details how this reduction could be achieved[2]. The bulk of the reduction is expected to come from efficiency and operations improvement in buildings, including lighting, insulation, and HVAC scheduling and optimization. Further reductions could be achieved by replacing single pane windows and replacing or improving steam lines around campus, but there are major legal, logistical, and sociopolitical challenges to these changes.

The remaining 17% emissions reduction is expected to come from off-site renewable energy. MIT sees it as imperative to invest in off-site renewable energy to meet its 2030 goal of 32% emissions reduction[2]. Although the electricity generated will not be used at MIT, it is enabling the construction of new renewable power, which does have a positive effect.

MIT expects 1% reduction to be driven by on-site renewable generation; put another way, it expects to generate 1% of its energy from on-site renewable energy. Currently, MIT generates just 0.01% of its energy from PV. Even in gloomy Boston, on MIT's small land space, PV could generate more than 25% of 2014 consumption from PV alone. A more realistic figure, assuming use of 5% of land area (which is less than available roof area) would put that figure at 1.5%, which would be a more substantial, while feasible, goal. Notably, Harvard has more than twenty times more PV than MIT, in the same neighborhood, and on comparable land area.

## Path to Scalable Carbon Neutrality for MIT

As described, only 15% of MIT's planned reduction from 2014 baseline emissions is scalable. That is, the purchase of off-site solar will not reduce the emissions of MIT directly. Perhaps an additional 25-50% reduction can be achieved through further efficiency improvements as detailed above. MIT cannot reduce its demand to zero, however, so it is imperative that it invests in local sustainable power generation. This generation needs to be safe, reliable, cost-effective, and have low GHG emissions. It will also be necessary, because of the dense energy demand of MIT, to draw electricity from a grid that is not fully renewable for the foreseeable future. One way to mitigate this action is by incorporating energy storage so that it can draw energy only when the grid is cleanest (e.g. when renewables peak). This storage also makes the supply more resilient and enables purchasing energy at the lowest cost, which is expected to coincide with lowest emissions. Harvard has started to adopt this policy as well[3].

An approach to achieve carbon neutrality by 2050, which greatly reduces emissions much sooner, is summarized in Figure 2. Starting from a baseline in 2014, demand is first projected to increase by 10% because of program growth. This growth jump is added every 20 years to account for anticipated growth. Next, in 2017 MIT's solar power purchase agreement (SPPA) in North Carolina went online and is being used to claim a 17% emission offset[4]. In 2020, MIT plans to bring online a new power plant, based on two 22 MW-e natural gas turbines for combined heat and power. Compared the current system, this is projected to reduce emissions by 10% compared to the 2014 baseline[5].

In MIT's CAP, it is suggested that by 2030, building efficiency improvements will reduce demand by 12-15%. There are several ways this could be accomplished, one of which is building scheduling. For example, unoccupied rooms can be controlled less actively, in a temperature range of 60-85 °F. Unoccupied consumption can be reduced by 30-80% using this and related tactics[6]. The cost of sensors, switches, and valves is expected to be more than offset by the reduced energy expense[7]. Another HVAC improvement that can be implemented is chilled water storage. Similar to UCF's approach, MIT can chill water when energy is cheap and clean, or use this method to balance fluctuating load on the Campus Utility Plant (CUP), allowing it to run at peak efficiency. Adding a tank that can store 1/3$^{rd}$ of MIT's current daily demand would cost approximately $4.5 million and consume 475 m$^2$ of space—potentially underground. As the campus becomes more efficient, this system is expected to store a day's worth of energy[8].

Notably, during this 35-year plan, the NE-ISO grid is required by the state of MA to become progressively cleaner, so it is important to shift toward grid energy over time. One major retrofit required to integrate efficiently with the grid is changing campus heating from steam to hot water. This will be a very expensive (order $100 million[9,10]) and time-consuming project because it is invasive—requiring plumbing/ventilation throughout campus. However, once the transition is complete, heat loss will immediately be mitigated, and combined with the above improvements, creates a path to 34% reduction in emission by 2030.

## Beyond 2030

Just as cold-water storage is proposed, hot water storage can also reduce our carbon footprint—with a similar cost and space requirement. By purchasing grid electricity when the price and emissions are low, MIT can store energy in hot water for later use. In the short term, this can enable CUP to operate at peak efficiency, and add resiliency in case of an outage on the order of 24 hours. Next, this plan proposes a deadline of 2035 to meet 1.5% of campus energy demands by solar PV at a cost of $500,000 [11]. This is a technically achievable task based on available roof area on campus and has already been done by MIT's neighbor, Harvard University[12].

The year 2040 represents a step change in the proposed plan. Here, MIT's CUP approaches end of life and is retired except for resiliency, and replaced by grid power. This enables to use of ground source heat pumps (GSHPs) to generate hot at a coefficient of performance (COP) above 4—that is, only 25% of the energy added/removed to building space by HVAC is consumed by electricity[13]. The total cost of replacing all boilers with GSHPs is estimated at $21 million[14]. This emission reduction is partially offset by the expiration of MIT's SPPA, which has a 25-year contact.

Two more improvements are employed to minimize demand before the final switch off the grid to wind and storage. First, heat recovery is implemented. Here, the exhaust gases in buildings with high air exchange rates are brought into indirect contact with incoming air in a counter flow heat exchanger to minimize the amount of thermal energy wasted by air exchange[15]. This method can, conservatively, reduce heating and cooling load by 25% at a cost of $2 million[16], resulting in a net emissions reduction of 4% compared to the 2014 baseline. This reduction is modest because the earlier switch of the HVAC system to hot water and GSHPs reduced the emissions greatly.

Importantly, by 2050, the grid is regulated to be 40% renewable. At this level of renewable penetration, it is assumed that during an average of 4 hours per day renewables have excess capacity. Here, the thermal energy storage of the hot and cold water systems for HVAC can be charged during these low carbon times (which will also be low cost)[3]. As these storage systems were designed to store a day of thermal energy and charge in 8 hours, they can get half of their energy carbon free—slashing emissions an additional 18%.

### Transition to a renewable micro grid

After completing the above demand reductions, which have a net effect of reducing average energy demand by nearly 50%, the final step of supplying this energy through a renewable microgrid is proposed. In the baseline case, this grid would consist only of offshore wind turbines collocated with StEnSea energy storage sphere off the coast of Boston. This energy storage technology, which was first proposed by Slocum[17], is described in the following Technology section—in short, it is based on storing energy in the pressure difference between a vessel and the water around it in the deep ocean.

Using hourly wind availability data from NOAA buoys[18] and hourly demand data from MIT Sustainability Data pool, for the entire year of 2016, the microgrid was sized. A cost minimization approach was used to determine the number of wind turbines, number of StEnSea units, and size of StEnSea pumps/turbines that could supply MIT without interruption for the 1-year period. Wind turbines were selected to have six MW-e rated power capacity, as this is a typical current size for offshore wind, although it is expected to grow over time to reduce cost. This power output is reached, based on Siemens specification, when the wind speed is 12.5 m/s, and is scaled with velocity to the third power below this wind speed[19]. Using the StEnSea unit size of 18 MWh-e of storage, the pump/turbine size was varied in order to select the number of hours of storage needed, but also be able to charge and discharge sufficiently fast. Notably, this resulted in being able to charge nearly four times faster than discharge, because of the need to oversize the wind farm to meet average load.

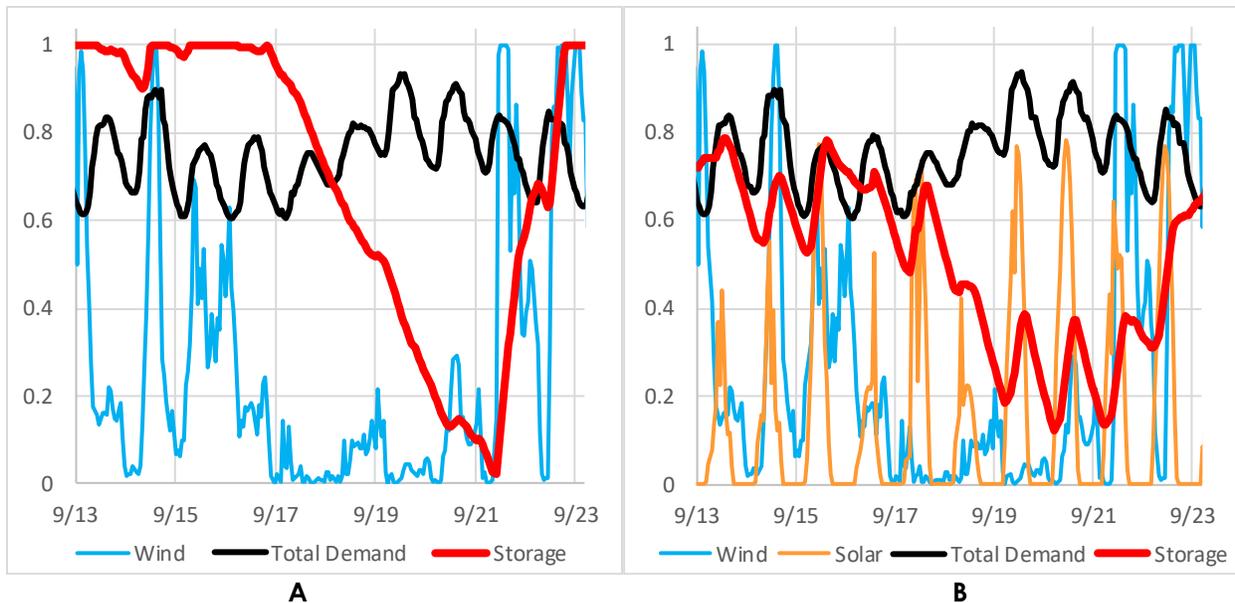

**Figure 1A:** Drastic need for storage capacity due to several days of low wind output, **B**: Effect of diversifying the microgrid by adding solar

This design of having a single source of variable supply results in the need for huge amounts of storage and power capacity oversupply to supply MIT with reliable power. The feature that has the most effect on cost and emissions is the summer period when wind output is low and demand is high—as shown in Figure 1A. Here, in a matter of 3 days, wind output is so low that the campus runs almost entirely from stored energy—resulting in the need to store 83 hours of energy. The second effect this has it to require significant power capacity over build because the average output over the summer months is low. As a result, the wind only approach requires 200 MW-e of wind and 2600 MWh-e of storage—or 6X MIT's average demand and more than 3 days of storage. Notably, just by diversifying to two wind farms off the coast of Boston, significant supply blending can occur, as shown in Figure 9. This comes at a cost of $1.4 billion dollars (including the cost of all campus upgrades and subtracting the avoided cost of purchasing energy from the grid). The raises the effective cost of energy to $0.30/kWh, up from the $0.06/kWh we pay today.

On the other hand, if the microgrid is made more diverse by adding the possibility of solar PV, cost and emissions decrease tremendously. Franklin county, 60 miles inland is the nearest location to MIT with reasonably low land cost to install a solar field, with a land cost near $50,000/acre[20]. The cost optimized microgrid in this case employs only 42 MW-e wind, 150 MW-e PV, and 1800 MWh-e storage. The resulting cost is only $400 million dollars, 70% less expensive than the wind only approach. In terms of LCOE, the cost is $0.13/kWh. This drastic cost decrease is possible because solar provides power when wind does not, replacing the nose dive with a saw tooth effect in Figure 1B. Although the cost per watt of solar is much lower than wind ($1.2/W vs. $4.9/W), a solar alone approach is actually even more expensive than wind alone, because solar has several days of low output in the winter. Diverse supply is key.

In the limit, with a diverse macrogrid supplied by variable renewables, costs approach or are even below today's energy costs[21]. In this case, based on a study on the nearby PJM market, only 4X generation capacity is required, compared to the 6X in the microgrid proposal. An even more drastic benefit arises in terms of hours of storage needed. In the diverse macrogrid case, only 22 hours of storage is needed. While this drives cost down another factor of 2, most importantly it reduces emissions by a factor of 3, as storage dominates emissions. Notably, although this option has the lowest cost and emissions, it is far beyond MIT's control to make this renewable grid a reality—but they can at least participate. Table 1 shows a comparison of these three carbon reduction approaches.

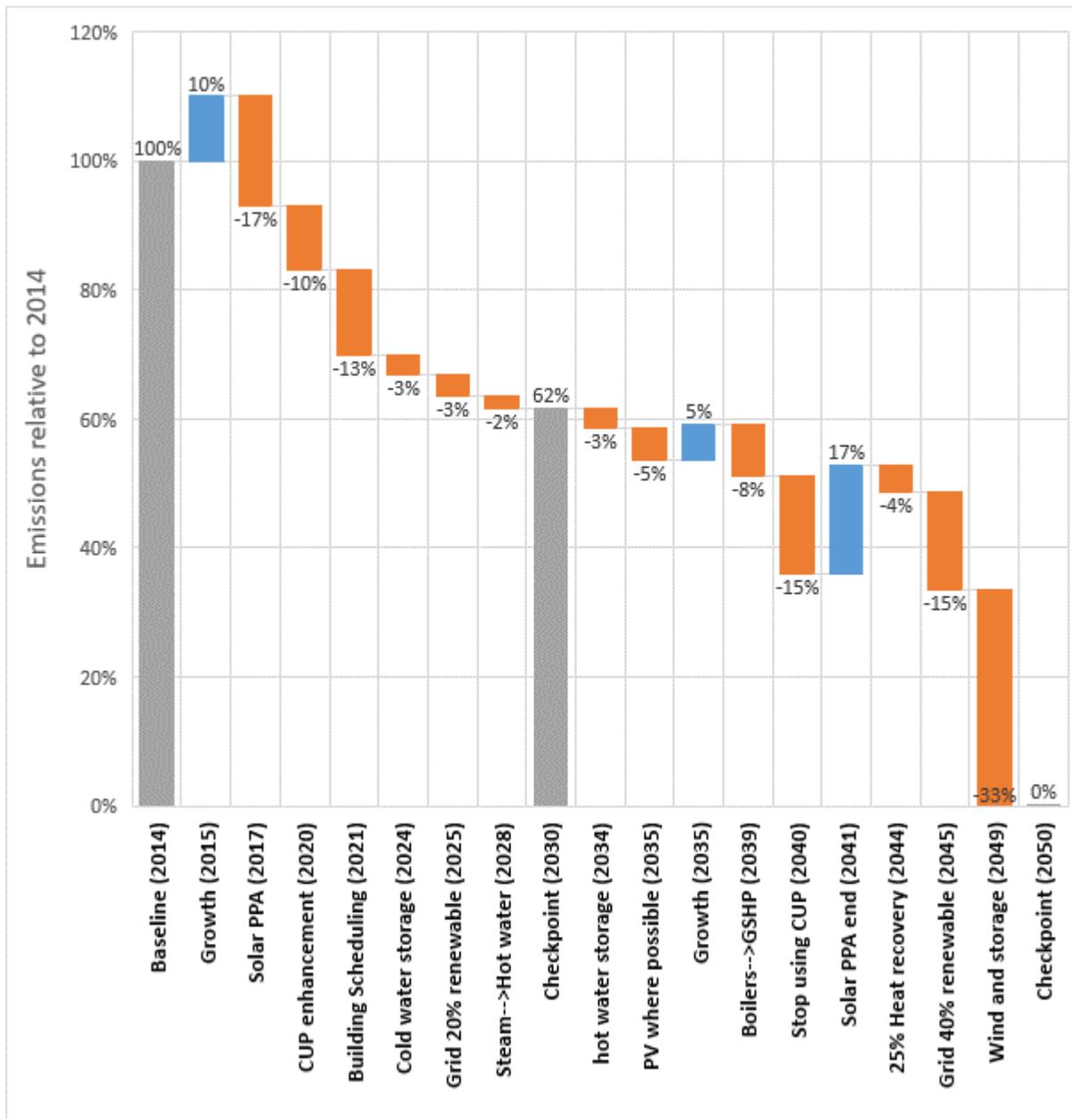

**Figure 2:** Summary of proposed MIT emissions elimination strategy

# Technology to Enable Scalable Carbon Neutrality for MIT

While most of the technology suggested for use is well developed, two important fields are still emerging—offshore wind and energy storage.

## Offshore Wind

The technical challenges underlying offshore wind revolve around mounting the turbines and connecting them to the grid. These challenges have current solutions, but further work is needed to reduce cost. Importantly, although the cost of an offshore wind installation is not expected to fall below onshore wind on a cost/turbine basis, the cost per energy may eventually be lower. This is made possible by the fact the offshore wind often has higher average wind speeds, and less variable wind speed. This enables a given turbine to have a higher power output and capacity factor, reducing the Levelized cost of energy (LCOE) and the amount of storage needed.

Offshore wind is a scalable energy supply, because there is enough area of high-speed wind available (4500TWh/year) off the coast off Massachusetts to supply more than 10x the energy demand of the state[22].

The carbon intensity of wind power, while very low compared to fossil generation, is important to account for. Compared to the current New England grid emissions of 320 kgCO2E/MWh, wind turbines emit about 15 kgCO2E/MWh assuming a capacity factor of 0.4 and a life of 20 years[19]. Importantly, carbon intensity is a strong function of capacity factor and curtailment. As will be seen, in a microgrid scenario without diverse supply, significant curtailment results from the need to have surplus supply to meet days, weeks, and seasons with low resource availability.

## Energy Storage

While there are many methods to store energy both currently available and under development, none have been proven to be geographically independent and low cost. For example, pumper hydro energy storage (PHES) has long had the lowest cost per energy (CPE) (neglecting highly limited compressed air energy storage (CAES)), but is not available/cost effective in most regions. On the other hand, electrochemical batteries including lead-acid and lithium based batteries are available, but at a much higher cost[23].

A novel energy storage approach first proposed in 2013 by Slocum[17] combines the approach of PHES with the more available pressure difference in large bodies of water—mainly oceans. In this concept, later termed storage in the sea (StEnSea), a concrete sphere of 30m diameter, located 700m below sea level, is proposed to storage energy by pumping water out of the sphere when excess power is available, and allowing it to fill later to generate power, as shown in Figure 3. High energy density is possible because of the high pressure deep below the ocean surface[24]. The cost of this form of energy storage on a CPE basis is 10-40% the projected future cost of utility scale lithium ion energy storage[23,25]. The lower end of this range is achieved when the number of hours of storage is high, which reduces the importance of the pump and turbine cost.

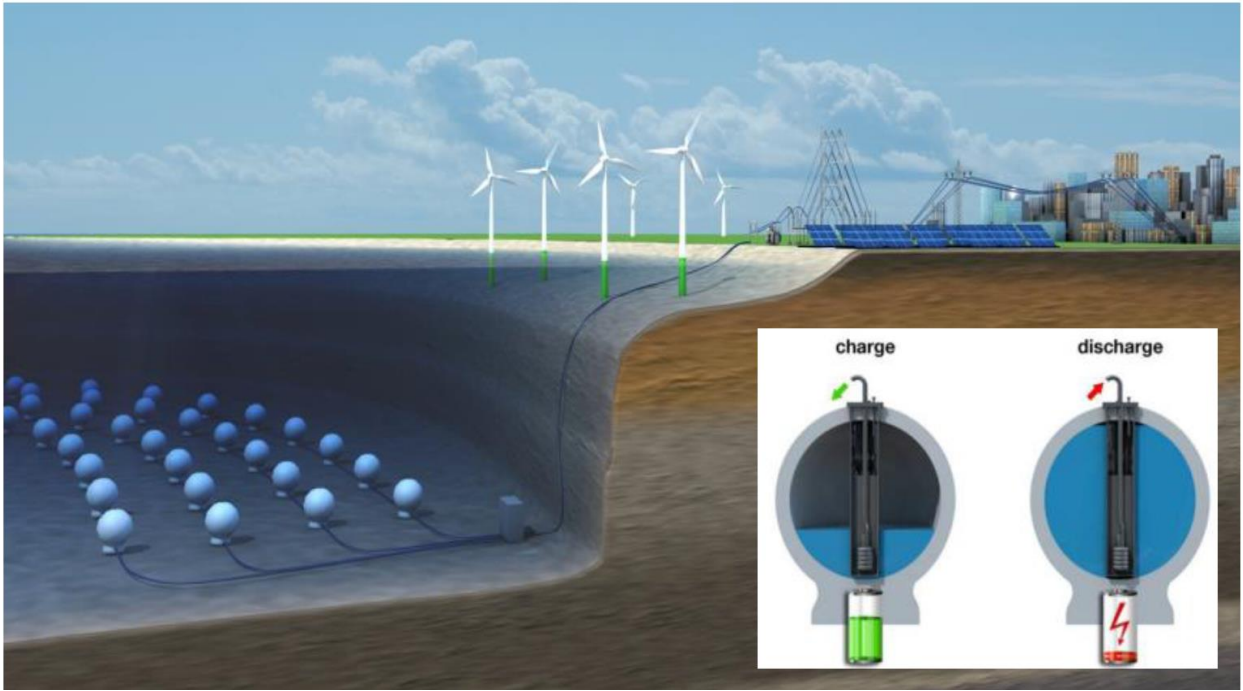

**Figure 3:** Storage in the sea (StEnSea) concept by Puchta

StEnSea is a scalable energy storage technology because it requires no rare earth elements such as lithium and cobalt, and is widely geographically available. For example, just 20 miles off the coast of Massachusetts, ocean depth is sufficient to store one day of Massachusetts energy in a 50 km$^2$ area, or 0.2% of Massachusetts land area.

The carbon intensity of energy storage can be significant, especially if many hours of storage are needed. The vast majority of the emissions from StEnSea are from the concrete[26] required, which results in emissions of 400 kgCO2E/KWh. This can be compared with emissions of 100 kgCO2E/kWh for lithium batteries[27], which last only ¼ as long and so have comparable emissions. Notably, these units are per capacity, and not per energy produced as is the case with grid and wind emissions discussed above.

To view the emissions cost of storage in that context, a system with 8 hours of storage can be envisioned. Here, the emissions of the wind plant would increase from 15 kgCO2E/MWh to 38 kgCO2E/MWh by adding 8 hours of storage. While this is a significant increase, the total emissions are still only 11% of the current New England grid emissions. Carbon intensity is a strong function of hours of storage required. As will be seen, in a microgrid scenario without diverse supply, many hours of storage are required.

## Cost Modeling

Current offshore wind costs are high compared to conventional generation, even though on shore wind has reached cost parity in many regions[28]. However, offshore wind power purchase agreements (PPAs) costs have been dropping rapidly for future projects, with contracts near $0.07/kWh for plants scheduled to come online in the mid-2020s. This cost is similar to conventional supply. For this report, these future PPAs are used to set the capital cost of wind at $5/W, and the turbines are assumed to last 20 years.

Detailed cost modelling has already been completed for StEnSea[25], so that applying it to MIT only requires projecting the life of the system and selecting the number of hours of storage. The authors provide enough information to decouple the cost per power (pumps and turbines) from the cost per energy (tanks). Here, the tanks are assumed to last 40 years based on prior results[29], but the turbines would be replaced after 20 years[30]. Then, the cost per power (CPP) and cost per energy (CPE) for 20-year life are $1.05/W and $72/kWh, respectively. Notably, this cost is very low compared to using lithium ion batteries for the task, especially because batteries cannot decouple CPP and CPE. In this case, the CPE for batteries is expected to be at least $400/kWh, assuming the batteries are replaced after 10 years.

The time value of money (TVM) is critically important to this analysis, especially when comparing solutions with different life. For example, because batteries last approximately half as long as StEnSea, they encounter a lower financing cost. This, fundamentally, is because instead of purchasing a 20-year plant on day 1, a 10-year plant is purchased. Then, 10 years later, the second half of capital expense (CAPEX) is required. For an interest rate, or IRR, of 7% this gives batteries a 25% benefit compared to purchasing two sets of batteries up front. Similarly, the fact that the StEnSea turbines are replaced after 20 years, compared to installing 2 sets up front, saves 38%—which is greater than the effect on batteries because of the longer time frame.

## Results and Conclusion

In this report, a plan for carbon neutrality for an off-grid MIT is presented. It is found that for MIT to isolate itself from the grid, a factor of 6X generation beyond average demand and 55-83 hours of storage is required. This massive over build results in 2-4X cost of energy increase and only marginal emissions reductions. In the best case, when solar and wind are employed, MIT's LCOE is ~$14/MWh and emissions are reduced by 75% from baseline. The biggest takeaway from this calculation is that MIT would be better off from a cost and emissions perspective to stay on the grid, assuming the current Massachusetts RPS is followed.

Another way to visualize this benefit is that in the wind only case, 63% of wind energy harnessed is curtailed, while in the wind and solar case (because only 20% of the wind turbines are needed) only 22% is curtailed. In the case of a diverse macrogrid, curtailment is expected to be minimal[21].

**Table 1:** Comparison of MIT energy supply and storage options

|  | Wind & Lithium | Wind & StEnSea | Wind/Solar & StEnSea | Renewable Grid | Units |
|---|---|---|---|---|---|
| Total Cost | $3,350,000,000 | $1,420,000,000 | $420,000,000 | $110,000,000 | $/MIT-20yr |
| LCOE | $0.58 | $0.31 | $0.13 | $0.08 | $/kWh |
| Cost per Carbon | $798 | $379 | $106 | $26 | $/MTCO2E |
| eROI | 1.63 | 1.08 | 1.52 | 3.04 | unitless |
| Annual Emissions | 26,399 | 26,968 | 18,238 | 7,197 | MTCO2E/yr |

When these four plans shown in Table 1 are evaluated on an energy invested verse recovered, or eROI basis, they are all low performance. In fact, in the base case of wind only[31,32], the eROI of 1.08 indicates that over the life of the system, only 8% more energy is recovered than invested. It is for this reason that energy must be sold at 4X the market rate. The best case is that of a renewable grid, but even here the eROI is only 3. In a truly optimized grid, eROI is expected to be 10 or greater, which is the case if lithium batteries replace StEnSea in the renewable grid scenario. While lithium batteries result in lower emissions and a higher eROI than StEnSea, their cost doubles the LCOE, even based on future cost of $300/kWh.

The cost per carbon, from a wind only approach is $379/MTCO2E. When wind is combined with solar location ~60 miles away in Franklin County, the cost drops to $106/MTCO2E. The most meaningful result of this effect is that diversifying variable renewable supply has a major effect of the amount of surplus supply and storage required. Thus, in the limit, connecting to a renewable grid is ideal. For example, if MIT stays connected to the grid, instead of investing in solar, wind, and electricity storage, it can reduce the cost per carbon to $26/MTCO2E.

While this plan is complex, it is science-based and each proposed reduction is the result of detailed calculations, referencing results previously achieved elsewhere. Thus, it is possible for MIT to eliminate ~90% of its greenhouse gas (GHG) emissions by 2050, although going it alone is not economically efficient. Many other implementations could be employed to reach this goal, as well. Some that are not explored here include modifying social behavior and implementing a self carbon tax. These approaches can be very effective, and efficient in addressing anthropogenic climate change. To eliminate the last 10% of emissions is more complicated, because even renewable sources have non-zero emissions. To reach actual carbon neutrality without offsets, some amount of carbon sequestration would be required.

Notably, scope 3 (indirect non energy) emissions are only narrowly addressed in the three universities' Climate Action Plans and in the proposed path to carbon neutrality for MIT. Some approaches to reduce scope 3 emissions include further subsidizing sustainable transport, banning the sale of beef and the meat of other ruminates because of the related methane emissions, and using MIT's technical strength to solve the challenge of carbon-free global transport. These approaches were not proposed here because they are difficult to model with any certainty, and they cannot lead to carbon neutrality, only modest reduction in emissions.

# Appendix

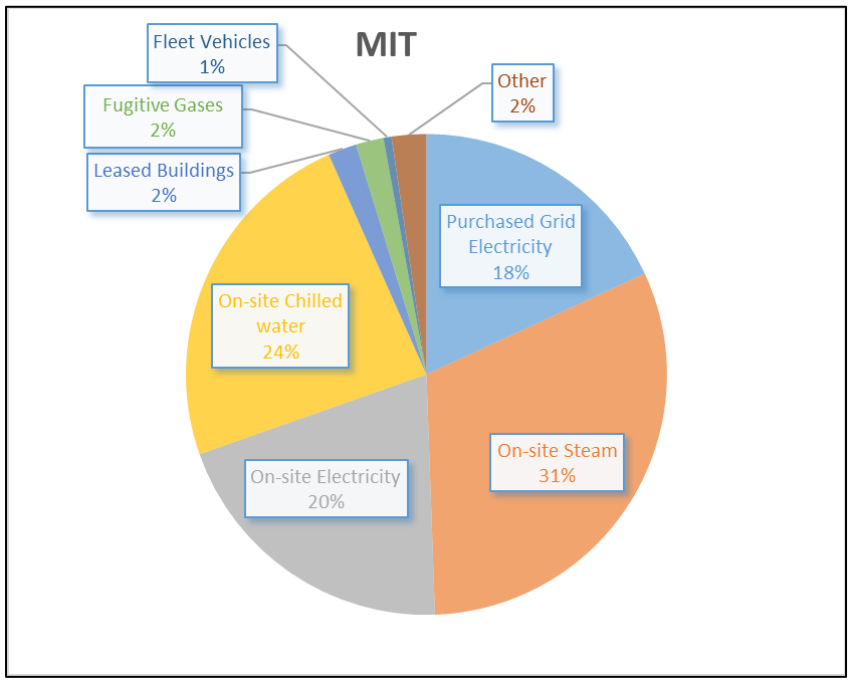

Figure 4: MIT 2014 emissions profile

| | Power | | | Emissions | | | | | | | |
|---|---|---|---|---|---|---|---|---|---|---|---|
| total | Elec (MW) | Cool (MW) | Heat (MW) | Elec | Heat | Cool | Year | Reduction | Method (for axis) | Cost | Space (m) |
| 63.05 | 22.95 | 6.15 | 33.95 | 100% | 100% | 100% | 2014 | 100% | Baseline | N/A | |
| 65.35 | 25.24 | 6.15 | 33.95 | 110% | 110% | 110% | 2015 | 10% | Growth | N/A | |
| 65.35 | 25.24 | 6.15 | 33.95 | 110% | 110% | 110% | 2017 | -17% | Solar PPA | negative | |
| 65.35 | 25.24 | 6.15 | 33.95 | 110% | 110% | 110% | 2020 | -10% | CUP enhancement | N/A | |
| 56.72 | 21.91 | 5.34 | 29.47 | 97% | 97% | 97% | 2021 | -13% | Building Scheduling | negative | |
| 56.72 | 21.91 | 5.34 | 29.47 | 97% | 94% | 94% | 2024 | -3% | Cold water storage | $ 4,625,717 | 474 |
| 56.72 | 21.91 | 5.34 | 29.47 | 97% | 94% | 94% | 2025 | -3% | Grid 20% renewable | N/A | |
| 56.16 | 21.91 | 5.34 | 28.91 | 97% | 92% | 94% | 2028 | -2% | Steam-->Hot water | $ 200,000,000 | |
| 56.16 | 21.91 | 5.34 | 28.91 | 97% | 92% | 94% | 2030 | 62% | Checkpoint | N/A | |
| 56.16 | 21.91 | 5.34 | 28.91 | 97% | 89% | 94% | 2034 | -3% | hot water storage | $ 6,433,161 | 659 |
| 55.72 | 21.47 | 5.34 | 28.91 | 97% | 89% | 94% | 2035 | -5% | PV where possible | $ 487,500 | |
| 58.72 | 22.62 | 5.63 | 30.47 | 102% | 94% | 99% | 2035 | 5% | Growth | N/A | |
| 5.63 | skip | 5.63 | skip | 102% | 55% | 99% | 2039 | -8% | Boilers-->GSHP | $ 12,995,283 | |
| 35.87 | 22.62 | 5.63 | 7.62 | 38% | 52% | 99% | 2040 | -15% | Stop using CUP | $ 11,584,443 | |
| 35.87 | 22.62 | 5.63 | 7.62 | 38% | 52% | 99% | 2041 | 17% | Solar PPA end | N/A | |
| 32.56 | 22.62 | 4.22 | 5.71 | 38% | 39% | 74% | 2044 | -4% | 25% Heat recovery | $ 2,050,000 | |
| 31.51 | 22.62 | 3.17 | 5.71 | 28% | 15% | 28% | 2045 | -15% | Grid 40% renewable | N/A | |
| 31.51 | 22.62 | 3.17 | 5.71 | 0% | 0% | 0% | 2049 | -33% | Wind and storage | | |
| 32 | 23 | 4 | 6 | 0% | 0% | 0% | 2050 | 0% | Checkpoint | N/A | |

**Figure 5: Raw calculation results**

|  | Power | | | Emissions | | |
| --- | --- | --- | --- | --- | --- | --- |
| total | Elec (MW) | Cool (MW) | Heat (MW) | Elec | Heat | Cool |
| =SUM(B3:D3) | =(172.4+28.6)*1000/24/3 | =617000/1.06/3600*10 | =1135000/1.06/3600*1 | 1 | 1 | 1 |
| =SUM(B4:D4) | =B3*1.1 | =C3 | =D3 | =E3+$J4 | =F3+$J4 | =G3+$J4 |
| =SUM(B5:D5) | =B4 | =C4 | =D4 | =E4 | =F4 | =G4 |
| =SUM(B6:D6) | =B5 | =C5 | =D5 | =E5 | =F5 | =G5 |
| =SUM(B7:D7) | =B6*(1+J7) | =C6*(1+$J$7) | =D6*(1+$J$7) | =E6+$J7 | =F6+$J7 | =G6+$J7 |
| =SUM(B8:D8) | =B7 | =C7 | =D7 | =E7 | =F7+$J8 | =G7+$J8 |
| =SUM(B9:D9) | =B8 | =C8 | =D8 | =E8 | =F8 | =G8 |
| =SUM(B10:D10) | =B9 | =C9 | =D9*(1+J10) | =E9 | =F9+J10 | =G9 |
| =SUM(B11:D11) | =B10 | =C10 | =D10 | =E10 | =F10 | =G10 |
| =SUM(B12:D12) | =B11 | =C11 | =D11 | =E11 | =F11+J12 | =G11 |
| =SUM(B13:D13) | =B12*0.98 | =C12 | =D12 | =E12 | =F12 | =G12 |
| =SUM(B14:D14) | =B13*(1+$J$14) | =C13*(1+$J$14) | =D13*(1+$J$14) | =E13+$J14 | =F13+$J14 | =G13+$J14 |
| =SUM(B15:D15) | skip | =C14 | skip | =E14 | 0.548400566268111 | =G14 |
| =SUM(B16:D16) | =B14 | =C15 | =D14/4 | 0.375173919407549 | 0.517119395266704 | =G15 |
| =SUM(B17:D17) | =B16 | =C16 | =D16 | =E16 | =F16 | =G16 |
| =SUM(B18:D18) | =B17 | =C17*0.75 | =D17*0.75 | =E17 | =F17*0.75 | =G17*0.75 |
| =SUM(B19:D19) | =B18 | =3.17 | =D18 | =E18*0.75 | =F18*0.75/2 | =G18*0.75/2 |
| =SUM(B20:D20) | =B19 | =C19 | =D19 | 0 | 0 | 0 |
| =ROUNDUP(A20,0) | =ROUNDUP(B20,0) | =ROUNDUP(C20,0) | =ROUNDUP(D20,0) | 0 | 0 | 0 |

| Year | Reduction | Method (for axis) | Cost | Space (m2) |
| --- | --- | --- | --- | --- |
| 2014 | 1 | Baseline | N/A | |
| 2015 | 0.1 | Growth | N/A | |
| 2017 | -0.17 | Solar PPA | negative | |
| 2020 | -0.1 | CUP enhancement | N/A | |
| 2021 | =-0.12*AVERAGE(E6:G6) | Building Scheduling | negative | |
| 2024 | -0.0300223061175371 | Cold water storage | =28*691000/1.06/365*0.333*10 | =691000/1.06/365*0.333 |
| 2025 | -0.0333480162610447 | Grid 20% renewable | N/A | |
| 2028 | -0.0189321385879879 | Steam-->Hot water | 200000000 | |
| 2030 | =SUM(J3:J10) | Checkpoint | N/A | |
| 2034 | -0.0285039948316245 | hot water storage | =28*961000/1.06/365*0.333*10 | =M8*L12/L8 |
| 2035 | -0.05 | PV where possible | =15*0.05*1000*1000*0.65 | |
| 2035 | =0.1*(SUM(J3:J13)-J11) | Growth | N/A | |
| 2039 | -0.0799004393211719 | Boilers-->GSHP | =58/1.06*1000000000/3600*0.5 | |
| 2040 | -0.152115392082112 | Stop using CUP | =(961000)/1.06*1000000000/36! | |
| 2041 | 0.17 | Solar PPA end | N/A | |
| 2044 | -0.0411046033737313 | 25% Heat recovery | =5000000*0.41 | |
| 2045 | -0.153142618686259 | Grid 40% renewable | N/A | |
| 2049 | =-SUM(J3:J19)+J11 | Wind and storage | | |
| 2050 | =SUM(J3:J20)-J11 | Checkpoint | N/A | |

**Figure 6: Raw formulas used to create waterfall emission reduction plot and costs**

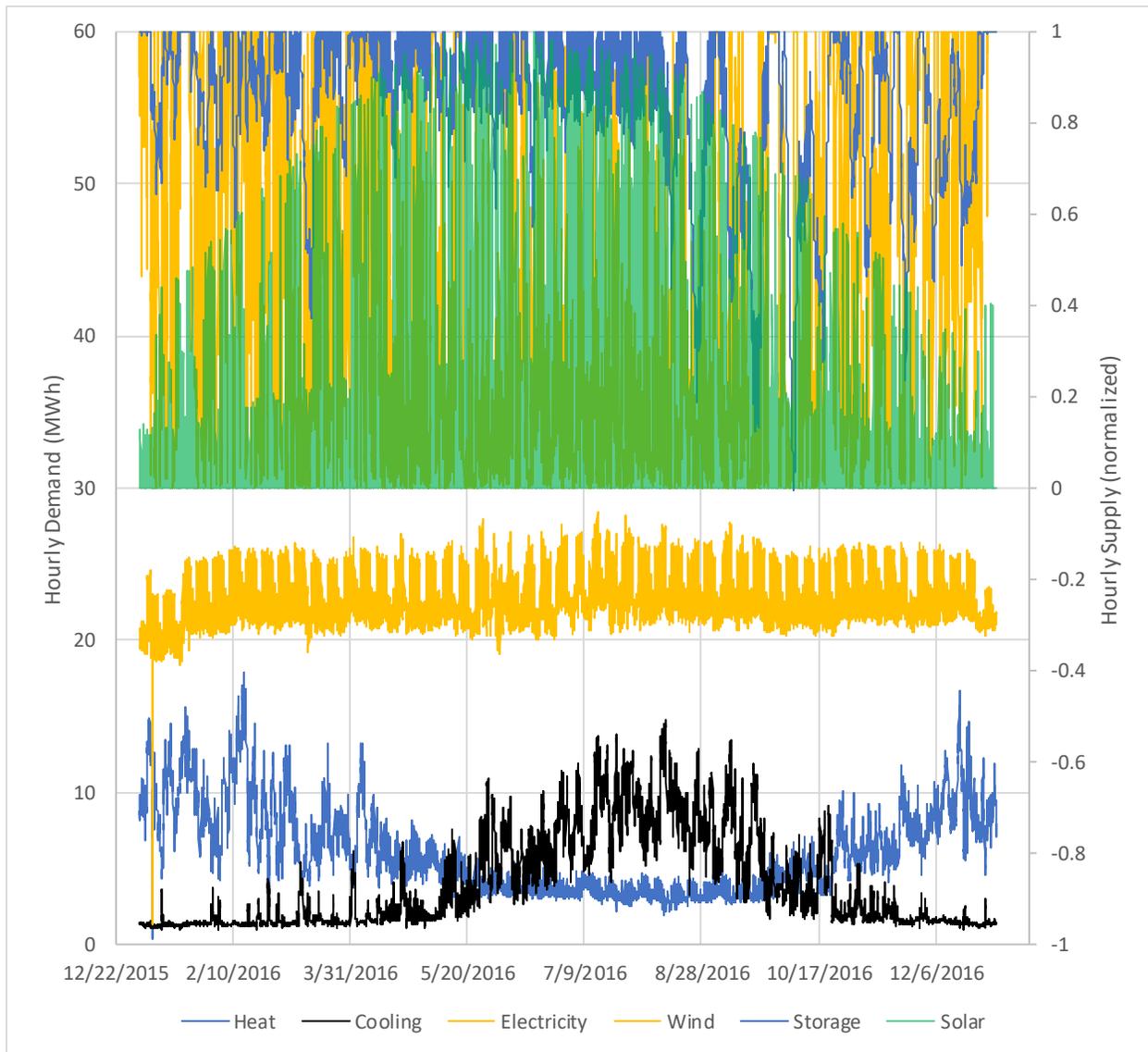

**Figure 7:** Result of 2016 MIT supply and demand if supplied by Wind and Solar plus Storage

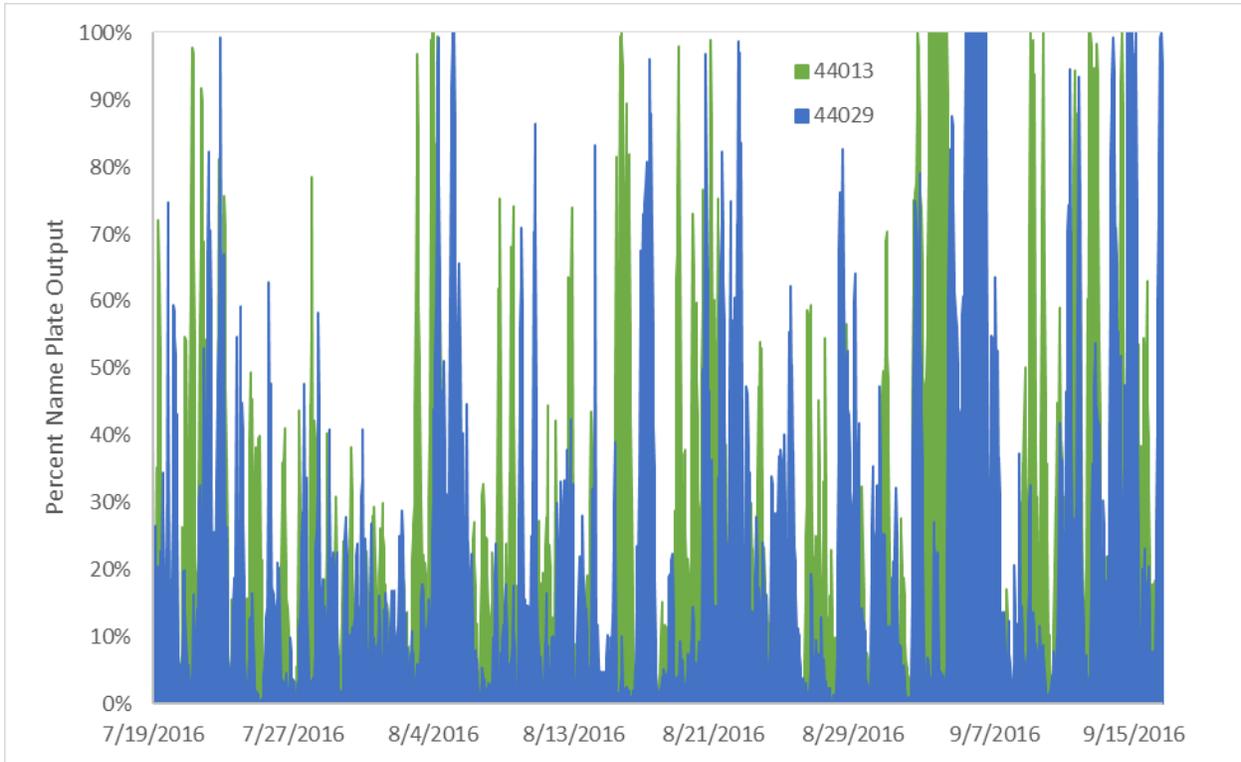

**Figure 8:** Result of having 2 wind farms ~50 miles apart off the cost of Boston

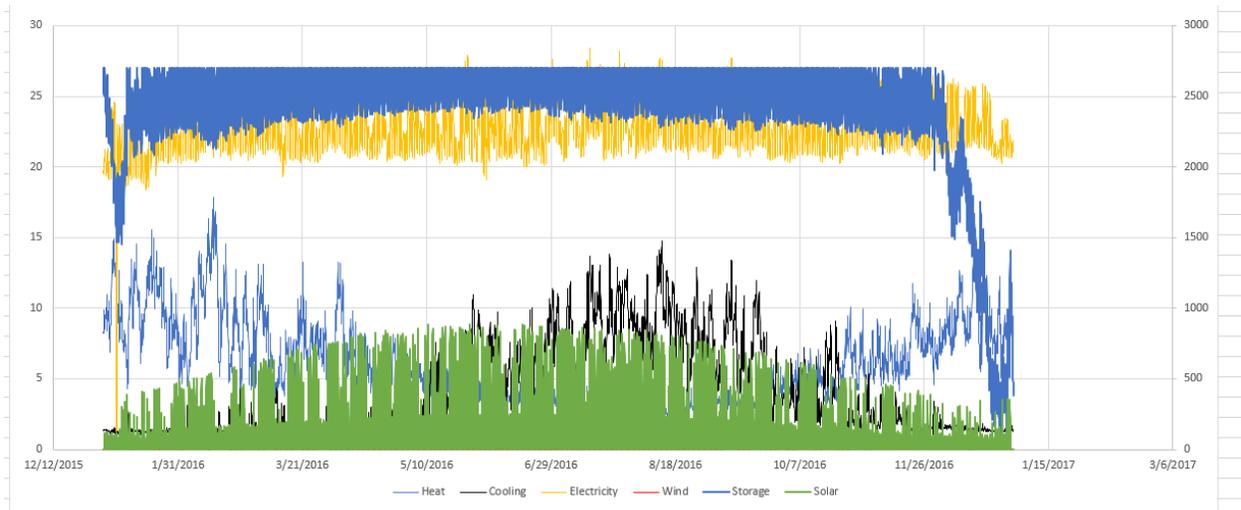

**Figure 9:** Result of Solar only case. Massive overbuild of capacity or storage is need to survive winter

# REACH PROJECT

## 2.S999 – Carbon Neutrality at MIT

**Julien Barber**
**April 2, 2018**

# POWER PURCHASE AGREEMENTS

To reduce greenhouse gas emissions, commercial, industrial and institutional (C&I) buyers have a few options. They can reduce demand-side energy loads, directly build on-site renewable generation, or use a power purchase agreement (PPA) to buy power and/or renewable credits from an off-site generation facility. Fundamentally, a PPA represents an energy contract between a buyer and a seller. The seller develops the renewable energy assets and sells the electricity and associated renewable energy certificates (RECs) to the buyer, often termed the 'offtaker'. The current paper reviews the current and future frameworks of PPAs, and analyzes the associated carbon reduction potential, costs, social issues, timeline, scalability, risk, uncertainty, and the land-use and lifecycle effects of renewable PPAs.

**How Power Purchase Agreements Work**
Two main contracts exist for C&I renewable PPAs – **physical** (or direct) and **financial** (or virtual/synthetic). The first, and most straightforward PPA is the physical/direct framework. In a physical PPA, the 'seller' directly delivers power to the 'buyer', who purchases the legal title to the energy produced and the associated renewable energy credits. In exchange for agreeing to purchase power from the seller over a fixed amount of time (usually 10 - 20 years), the buyer locks in a fixed cost of electricity (COE) [1]. In doing so, the buyer is able to insulate themselves with a stable COE in a volatile energy market. Ensuring stable prices over time is of particular interest for C&I customers who have high electrical operating costs due to a large electrical demand.

The seller typically develops, owns, and operates the renewable energy project and delivers the output electricity to a specified delivery location. From the delivery location, the buyer takes ownership of the electricity and RECs and is responsible for delivering the power to its target load. Since the purchased renewable energy is inherently intermittent, a physical PPA often includes scheduling and balancing services to ensure that all energy requirements are met [1]. However, since the power is directly delivered to the buyer, the physical/direct PPA must be made between two parties that share the same electric grid in a *deregulated* retail energy market. Deregulated energy markets - also known as Regional Transmission Organizations (RTOs) or Independent System Operators (ISOs) - are competitive, organized markets where the power that is generated enters the wholesale market and is traded like any other commodity. The interconnectivity of the deregulated grid allows electricity trading across geographic locations and drives the competitive wholesale price of electricity. Therefore, if a 'buyer' wants to implement a physical PPA, their source of demand must be located in a *deregulated* energy market that shares an energy grid with the renewable generation facility. If an organization is in a *regulated* energy market, they **cannot** enter into a physical PPA with a seller. Regulatory constraints limit the availability of physical PPAs to some buyers. Moreover, the physical delivery of power to C&I buyers is often impractical and/or expensive due to resource availability, physical space requirements, or inadequate transmission capability [2]. To get around the issues associated with regulated markets, resources availability, and infrastructure, C&I customers have begun to use **financial**/virtual/synthetic PPAs.

Similar to the physical PPA, the **financial** PPA is a contract between two parties (buyer and seller). However, in a financial PPA, no power is directly delivered to the buyer. Rather, the seller delivers the renewable electricity directly to the grid, receiving the wholesale market price. In the negotiation of the financial PPA, the buyer and seller agree on a fixed rate known as a 'strike price'. For the agreed upon price, the buyer receives the RECs and is payed (or charged) the difference between the market rate and the annual strike price. If the market price is higher than the strike price, the buyer is paid the difference. If the market price is lower, the buyer is charged. In this manner, C&I buyers in regulated markets, or in geographies with low renewable resources, can advance their sustainability objectives, while continuing to reduce and stabilize their long-term energy costs. A summary of a physical and financial power purchase agreement is shown in Fig. 1 [3].

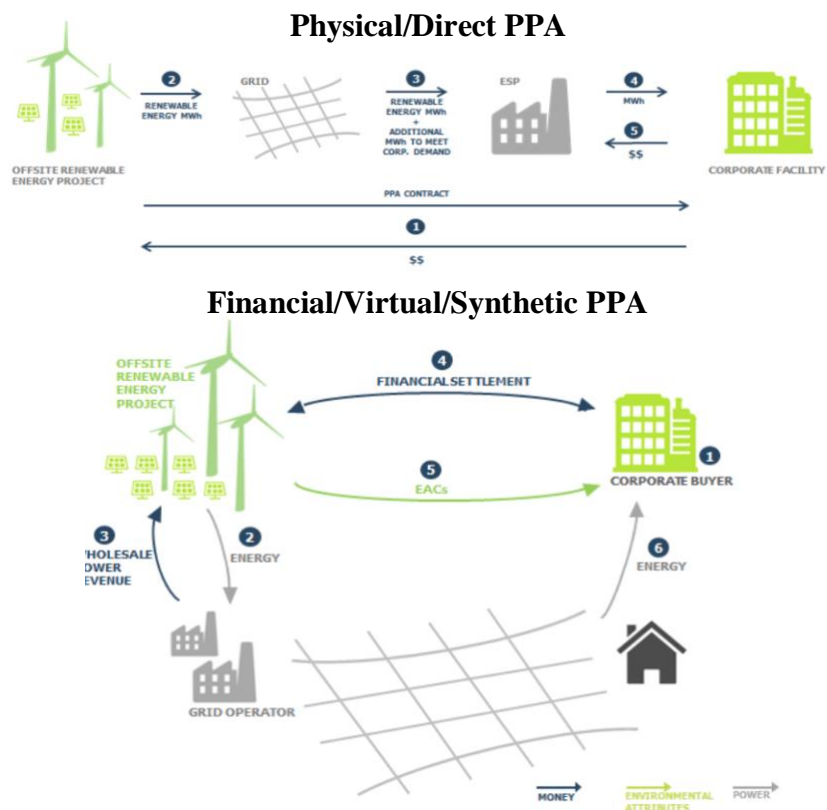

*Figure 1 - Working principles of physical and financial power purchase agreements (PPAs) [3].*

**Why Choose a Renewable PPA**
The PPA was created to encourage investment in renewable energy projects. By fixing the cost of electricity over 10-20 years, the developer is able to secure funding due to a reliable source of income, and the buyer locks in a stable cost to insulate against volatile energy prices. For C&I buyers, the stable PPA energy price allows for predictable budget management, and normally generates positive financial savings due to rising retail energy prices. Moreover, a renewable PPA allows buyers to reach their internal and/or state mandated renewable energy targets through either the direct purchase of power, or indirectly through RECs [4]. This can be an attractive options, especially when resource and land availability does not permit the development of on-site renewable energy.

PPAs are structured in a manner which simplifies the purchasing experience and encourages adoption. Perhaps the most attractive benefit to buyers is the arrangement of no, or low upfront costs. By limiting the upfront costs, a customer is only responsible for the contracted terms of agreement and only pay for what is generated. Since the seller develops, sizes, and installs the generating facility, they also take care of all operational and maintenance costs associated with the project. By owning the physical equipment, the seller alleviates the system performance risk from the buyer, and removes any barriers to entry associated with complicated systems design and permitting processes. Under this framework, the buyer is simply responsible for a utility bill, while the operation and performance of the system is cared for by the seller [4].

Conversely, in the long run, a buyer will pay less and save more by purchasing their own renewable energy facility. Leasing a project may also make more sense due to shorter contract terms (7-10 years) and lower costs of capital since the buyer is now responsible for the operation and maintenance of the system. By entering into a PPA, the buyer is foregoing all control over system design and performance. By purchasing or leasing generation capacity, buyers can retain control over their operations – however at an price of increased cost, risk, and greater operational complexity.

**Adoption of Renewable PPAs for C&I Buyers**
According to the Rocky Mountain Institute, traction for C&I PPAs first began in 2012 and has grown rapidly ever since, now exceeding 50% of the total PPA market [5]. In 2013, 0.77 GW of renewable corporate PPAs were signed, growing to 3.26 in 2015 and reaching 3.11 GW in 2017. Fig.2 shows the aggregate growth of PPA deals in the C&I sector from 2013-2018 (YTD).

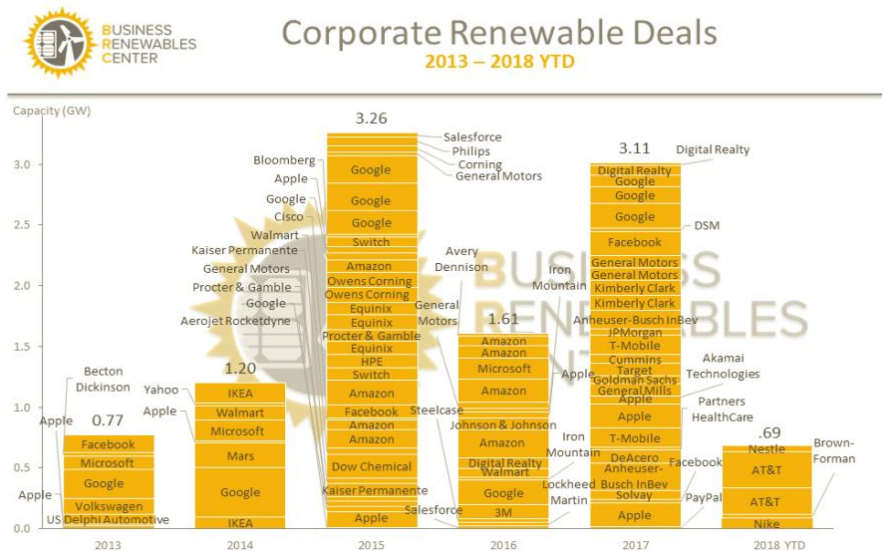

*Figure 2 - corporate renewable power purchase agreement deals from 2013-2018 (YTD) [6].*

A select collection of companies make up a significant market share of C&I PPAs, and have paved the way for other companies to follow suit. In 2018, 128 of some of the world's most influential companies – including Google, Apple, Coke, Nike, Goldman Sachs, IKEA, Visa, Starbucks, Microsoft, and Facebook – have signed the RE100 agreement to source 100% of their operational energy use from renewable energy. In 2015, the total energy use of the RE100 was approximately 146 TWh – roughly the electrical demand of Poland or the state of New York [7]. To reach these goals, most companies are turning to renewable energy PPAs as well as direct on-site generation where possible.

As an example, Google is using wind to reach their renewable energy targets. The company is able to purchase a **physical** PPA that directly buys energy from an off-site renewable energy developer on the same grid as their data facilities. The power is generated and sold into the grid at the local, wholesale market price. Google obtains and retires the RECs generated from the facility to count towards the energy demand from their data centers. Since, in this particular case, the retail grid price is less than the set PPA price [8], Google takes a slight financial loss on operating costs. However, in exchange for this cost they are able to claim the renewable credits, and any future cost savings if retail electricity prices rise.

A key component to C&I customers interested in PPAs is the idea of 'additionality'. Essentially, additionality means that a PPA must result in the development of renewable energy that adds additional carbon-free capacity to the electric grid. In an indirect way, additionality can also be considered true when actions result in the increase in demand for renewable energy through market pressures that encourage new investment [8]. However, RECs on their own are not enough to provide the stable and sizable cash flows that renewable energy developers need to finance their new (or additional) projects. Through agreeing to the terms of a PPA, C&I customers are stating they will buy all energy produced from the generation facility for the lifetime of the contract. Doing so transfers the power price risk from the project owner to the customer signing the PPA. With a creditable company (such as Google), the developer literally takes the power purchase agreement to the bank to obtain financing that allows new renewable generation to be built into the grid. Simply buying RECs would not provide the long-term purchasing agreement that the bank would require to fund the new, additional renewable energy project.

**Alternative Financing Schemes**
The purpose of a PPA is to provide sufficient financial backing to allow developers to add additional renewable energy capacity to the grid. The renewable energy industry needs financial innovations, such as the PPA, to expand the availability and lower the barrier to entry of renewable energy projects. New, alternative financing mechanisms outside of PPAs include public market capital, hybrid bond financing, and financing through the sale of green attributes.

Public capital vehicles set up a pool of cash flows that create low-risk, easily tradable commodities that can attract a wide range of investors. By pooling assets and selling off ownership interests, the public capital vehicle allows diverse investment, marketability of the security, and transparency of the market price through public trading [9]. Together, this creates a low risk investment opportunity that can attract capital at a lower rate of return. Providing a risk-averse rate of return on investment allows the developer sell electricity at a lower cost and become competitive in the energy market. A NREL technical report found that the use of asset-backed securities (ABSs) for capital incorporated as debt, and master limited partnership (MLPs) for capital incorporated as equity, can reduce a project's levelized cost of electricity (LCOE) by 8%-16% compared to a typical PPA [9].

Hybrid bonds are another recent financing mechanism that combines a *portfolio* of renewable projects. In a generic bond, the issuer makes scheduled payments to the investor in return for upfront investment capital. The hybrid bond for renewable energy financing functions in a similar manner, where in exchange for investment capital, the developer pays a rate of return on the hybrid bond. The cash flow from future revenues from the *pooled* renewable energy projects are securitized and form the basis of the returns for the hybrid bond [10]. The act of pooling renewable energy projects leads to securitization of the cash flows and results in lower market and credit risk.

Finally, financing through carbon credit markets, which include renewable energy credits (RECs), is an additional method to generate capital for renewable investment. Similar to the previous financing mechanisms, a carbon bond is created that securitizes future revenue from the REC to raise initial capital from investors. The developer sells the carbon bond to the investor who helps fund the initial capital cost of the project. In return, the investor is entitled to collect a repayment backed by the sale of the RECs every year until repayment, with appropriate interest.

Despite the variety of financing models, PPAs remain the most widely adopted financing mechanism for renewable energy projects. These contracts offer buyers tangible means to reduce carbon emissions while generating new renewable capacity. However, by purchasing a PPA, the buyer claims reduction in their personal emissions while also reducing the average carbon intensity of the grid in which the developer installed the facility. Therefore, double counting of emissions exaggerate the true effect of PPAs. The topic of double counting will be discussed in a later section. For now, we will assume the REC accounting scheme is valid, and that the renewable PPA is bundled with RECs that are certified additional.

**Cost & Carbon Reductions**

The price of renewable PPAs vary greatly depending on contract type (direct or financial) and on the wholesale and retail market prices in the region being negotiated. In particular, virtual/financial PPAs often enter into a contract-of-differences in which a strike price is set. The strike price is often kept confidential, and as such there is limited data for financial PPAs. However, the available, nationally averaged levelized prices of wind and solar PPAs that bundle both the sale of electricity and RECs are shown in Fig. 3. In 2016, the average cost of wind and solar are approximately the same at ~$25/MWh. To illustrate both the financial and direct PPAs, two cases will be considered.

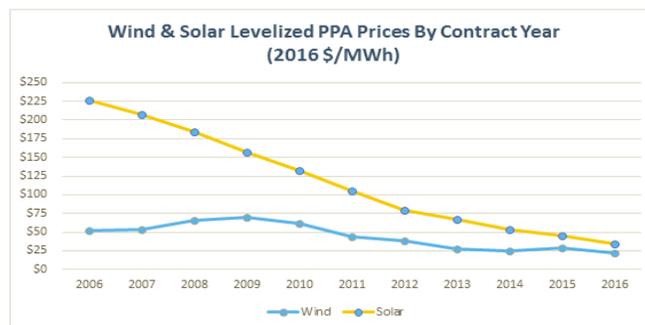

*Figure 3 - Prices of solar and wind PPAs, levelized over the full contract term (Source: Lawrence Berkeley National Laboratory)*

First, the direct PPA example considers *additional* wind bundled with RECs purchased at $25/MWh and installed in the ISO-New England grid. By purchasing a direct PPA, MIT is essentially "removing" its electricity demand from the emissions accounting of the region since it owns legal right to the power and RECs. In doing so, the contribution of the installed renewable capacity due to the privately funded PPA should *not* count towards the carbon intensity of ISO-NE grid to avoid double counting. To offset the entire MIT electricity demand purchased from the grid in 2017 (177,725 MWh) at a current grid intensity of 0.330 MT $CO_2$/MWh, MIT would need to pay roughly $4.44M annually to reduce emissions by 60,249 MT $CO_2$ (or 28% of their 2014 total). Conversely, if the campus were 100% electric, MIT would need to spend ~$21.04M annually to claim 100% renewable energy. Essentially, MIT would need to pay $74/MT $CO_2$ due to the carbon intensity of the grid and PPA price per MWh.

Next, let's consider an offsite virtual PPA (VPPA), such as the Summit Farms Case in North Carolina. The VPPA is slightly different in that energy is not directly supplied to the ISO-NE grid where MIT draws its power. Therefore, there are two carbon accounting scenarios to be considered. First, if we simply apply the RECs generated at the offsite location to MIT's energy demand then the result is the same as the direct PPA. By setting the strike price of the VPPA to result in an average levelized cost of energy of $25/MWh, MIT will need to pay the same annual cost as in the direct PPA to "offset" their emissions ($4.44M for 28% and $21.04M for 100%). However, if we draw our boundaries around the ISO-NE and PJM grids, MIT's investment in additional renewable capacity in North Carolina actually offsets the equivalent load from the PJM grid with a carbon intensity of 0.450 MT $CO_2$/MWh. Therefore, the $4.44M investment would now offset 79,976 MT $CO_2$ (or 37% of MIT's 2014 baseline), resulting in a cost of $55/MT $CO_2$. Emissions accounting based on unbundled RECs is referred to as a "market-based" accounting. RECs permit MIT to take ownership of the environmental attributes associated with the renewable energy from the off-site generation facility. However, the claim that all RECs are equivalent to direct emissions offsets is a topic of hot debate [12-14].

**Accounting Issues: Political, Social, and Ethical Concerns**
Emissions accounting becomes complicated with PPAs as it depends where you choose to draw your boundaries. On a national level, installing *additional* solar in North Carolina is a smart investment since you can reduce national emissions for a lower cost. However, if we simply look at the state of Massachusetts, the investment in North Carolina has done nothing for reducing emissions within our own state, nor arguably for MIT. With borders drawn around the ISO-NE grid, the carbon emission intensity remains unchanged. What has changed is the emissions around the PJM grid, even though MIT has legal claim to the environmental attributes. The accounting becomes even more abstract if the emissions reductions gained from a VPPA, such as Summit Farms, are distributed within the regional electricity grid and shared by all local residents. This leads to double-counting of emissions reductions by MIT (the REC holder) and PJM the grid operator. The market-based accounting scheme attempts to address this by assigning the green credit exclusively to the owners of the RECs. However, if the accounting were to be followed, the addition of renewable capacity should be excluded from the PJM carbon intensity calculation. However, currently PJM does not exclude renewable capacity from external PPAs in their grid mix, since that power is being used to satisfy local demand. Consequently, there are complex inter-regional policy frameworks that must be agreed upon in order to ensure accurate emissions accounting [14].
In addition to double-counting and political claims to emissions reductions, there are social concern that RECs don't accurately attribute the true environmental benefits of 1 MWh of renewable energy. As mentioned previously, equating RECs to carbon offsets requires one to prove the all of the electricity used to generate the RECs was made possible due to the purchase of the bundled PPA and RECs - i.e. the project was "additional". Assuming additionality is true, quantifying the precise environmental benefit produced by using renewables to offset fossil-based generation is difficult. Factors such as time of use, location of generation in the grid, and the source of energy being offset can alter the environmental benefits due to 1 MWh of renewable generation [15].

Finally, emissions accounting based on RECs develop certain ethical concerns that make fool-proof accounting schemes difficult to develop. For example, VPPAs will be used to decarbonize the grids that have the highest potential renewable resources first, since that will be the cheapest option. If the green credits are passed along to external buyers, what happens when the entire grid is decarbonized? The REC owners will claim the environmental attributes, but the residents who own no renewable credit, will not be physically emitting any emissions. Under the REC accounting scheme, this will force the current residents to purchase RECS in regions with lower resource availability resulting in higher costs. If the region is already supplied with low-cost renewable power, strict enforcement laws would need to be in place to ensure energy users purchased these more expensive RECs. Doing so will not only be politically difficult, but will also marginalize regions that have less financial resources available to adopt renewables at the early stages of its development.

A fair amount of literature analyzes the issues associate with RECs and PPAs. However, due to the inherent ambiguity of emissions accounting, there are few concrete recommendations to improve the current system. Location-based accounting schemes reduce the incentive for C&I buyers to invest in PPAs. Since the cost benefits of renewable PPAs are already marginal, reducing the ability to claim impact through a VPPA negatively impacts the development of new renewable projects. Conversely, market-based accounting creates more uncertainty through buyers claiming any and all types of RECs as equivalent to renewable offsets. RECs purchased in the voluntary market and/or from non-additional development are **not** equivalent to offsets and therefore allow for misrepresentation on top of double-counting. Consequently, there is currently no agreement on an accounting solutions. The World Resource Institute, Carbon Disclosure Project, and World Business Council for Sustainable Development all advocate for a dual reporting scheme which allows reporters to claim environmental attributes due to RECs. These reporting standards are under scrutiny by some scholars, stating that only locational grid average accounting should be used to publicly disclose emissions [15] and shared accounting schemes [16] should be used. However, these sources recognize that neither is perfect, and don't provide supporting detail on execution of these plans. Until inter-state regulations can properly account for emission reductions, a buyer may be satisfied knowing they have locked in stable energy rates and contributed additional green energy to the national grid. Undoubtedly there are accounting issues that each institution should evaluate and consider before engaging in a PPA to reduce Scope 2 emissions.

**Timeline and Scalability**
Despite the accounting ambiguity, renewable PPAs are definitively increasing the amount of renewable generation on the grid and decreasing national emissions. More than 40% of Fortune 500 companies and over 500 schools across the US have targets relating to renewable energy procurement [17]. For universities like MIT, renewable PPAs offer a scalable financing mechanism for funding additional renewable energy projects that can contribute to reducing emissions, outside of energy efficiency and on-site resources. However,

C&I PPAs have begun to gain traction and show no sign of slowing down. Key environmental, economic, financing, and market drivers have many analysts predicting a drastic uptake in renewable PPAs as corporate buyers seek to increase their renewable energy procurement and improve brand image [18]. Granted, commercial and regulatory hurdles, as well as ethical and social concerns will continue to make it difficult for corporate PPAs to be deployed. However, as demand for renewables increases, targeted government interventions to support the innovative accounting schemes and financing structures discussed previously are likely to further accelerate adoption of PPAs and foster growth of the renewable industry.

**Risk and Uncertainty**
However, to foster the rapid growth renewable PPAs, there are a host of risks that must be mitigated for corporate or institutional buyers. Fortunately, to accommodate the various acceptable risk profiles of the many potential buyers, PPAs are able to apply a wide variety of contracting and pricing arrangements. In the current section three broadly defined categories of risk management will be discussed.

First is market and regulatory risk. In an energy market, all users are exposed to the volatile shifts in supply and demand associated with market and regulatory fluctuations. Volatile supply and demand leads to increased risk of volatile energy prices. For large energy consumers, this price volatility may become unacceptable. Renewable PPAs can help hedge this risk because once developed, renewable generation has a stable operating cost, which results in predictable, long-term energy rates. In proper due diligence, it is essential for a buyer to understand the market and regulatory factors, scheduled generation facility retirements, new generation construction, facility load profile, transmission planning, and past and future policy decisions to ensure a profitable contract can be set [19].

The second risk profile relates to project execution and operation. There are many different renewable developers that are looking for creditworthy C&I buyers to help get their projects off the ground. However, not all developers are created equal. As the PPA market grows, buyers must be sure to develop contracts with trustworthy, creditable developers to ensure the project develops to specifications and on schedule. Failing to enlist reputable developers increases the chance of underestimated construction costs, delays in project completion, failure to meet performance targets, prolonged generation outages, and environmental incidents – all of which can affect the buyer's bottom line. Performing due diligence on developers and properly managing the initial PPA contracting can help mitigate these risks. For example, requiring developers to post a certain amount of credit to represent their confidence in the project and agreed upon PPA can help insulate the buyer from project failure or underperformance. Fortunately, reputable companies who have been using renewable PPAs – such as Google and Amazon – have shared their procedure and best practices to help other companies successfully navigate these challenges [20].

The third category to consider is environmental and reputational risk. C&I buyers around the world are working to mitigate their environmental impact and improve their public image through investing in renewable energy, green certificates, and implementing sustainability and climate related goals. Currently, the public has shown support to companies and institutions that have shown resolve in combating climate change. However, not all claims to credits are deserving. As discussed previously, the purchase of non-additional RECs are not equivalent to direct offsets, and are at best a small subsidy for renewable projects. Buyers who are serious about their climate goals and want to continue to foster trust and transparency with the public have to be careful not to claim credit where credit is not due.

**Summary**
Direct and virtual PPAs present MIT and other C&I buyers with a financing scheme that can stabilize electricity costs, produce a positive return of investment, and serve to add additional renewable energy to the national grid. Placed in the ISO-NE grid, a renewable PPA is estimated to cost MIT approximately \$74/MT $CO_2$. Moreover, as the price of renewable continues to decline, the financials of PPAs will only improve. With falling contract prices and 3.11 GW of corporate PPAs signed in 2017, C&I PPAs are quickly becoming an attractive option for meeting emissions reduction targets. However, issues relating to double-counting, inter-grid communication, misrepresenting RECs as offsets, and future ethical concerns all present barriers for successful, long-term implementation of renewable PPAs. Significant government interventions to support the innovative accounting schemes and financing structures outlined in this report will be required to ensure the environmental attributes and financial rewards of PPAs are properly placed.

High-temperature phase-change thermal energy storage
Reach Project Report – 2.S999
Colin Kelsall
4/2/2018

**Introduction:**

This reach project proposal continues exploration into energy storage options for MIT and investigates an experimental technology capable of the high energy densities required for urban storage installations. Due to the limited availability of traditional renewable energy resources on the MIT campus, local production of electricity and heat will continue to be a carbon-intensive process. In the long-term, I see no alternative to converting MIT to grid electricity for all our energy needs and removing the carbon emitting CUP on campus. This approach, however, only makes sense if the carbon intensity of the grid is less than that of MIT's CUP, something that may not happen for many years. By storing energy produced off peak by green energy sources, MIT can decrease the relative carbon intensity of electricity purchased from the grid. Assuming no decrease in nuclear capacity, the grid is projected to be carbon-free by 2069, but MIT could be using carbon-free electricity by as early as 2048. To take maximum advantage of this effect, MIT should be able to store its full daily energy consumption (heating, cooling, and electricity) on campus. In order to economically store this energy requirement, ~2000 MWh, an energy storage medium with very high power density is required. Lithium-ion batteries, the gold standard for electrical energy storage, would be able to store 2000 $MWh_e$ in approximately 5000 $m^3$ at a cost of $500 million excluding installation costs. This report will propose an experimental high-temperature phase change thermal storage system with potential to store 2000 $MWh_e$ in roughly 4000 $m^3$ at a cost of $80 million excluding installation costs.

**Proposed Technology/Literature Review:**

A significant portion of MIT's energy consumption in 2017 was in the form of steam and chilled water, 44% and 28% respectively [1]. In my last report, I proposed the integration of phase-change thermal energy storage media (PC-TES) into the campus CUP system to allow flexibility in when MIT buys power from the grid. Although the concept of energy storage remains instrumental to this report's approach to carbon neutrality, an alternative to the previously proposed PC-TES materials have been investigated. Traditional phase-change thermal storage mediums operate at the relatively low temperatures required for building heating and cooling (~80°C and ~8°C, respectively), but materials with higher melting temperatures typically have higher latent heats of fusion and thereby high energy densities. Silicon [Si] ($T_{melt}$ = 1414°C) and alumina [$Al_2O_3$] ($T_{melt}$ = 2072°C) are two such materials with volumetric, phase change, thermal energy densities of 1283 $kWh_t/m^3$ and 1119 $kWh_t/m^3$, respectively. Additionally, these abundant bulk materials are very cheap at $3/$kWh_t$ and $1.40/$kWh_t$ of thermal storage capacity, respectively. Energy stored as heat in these high-temperature materials can be converted back to electricity at a high conversion efficiency when compared to lower temperatures. At 1414°C, Carnot efficiency (theoretical maximum thermal efficiency) is 82% and there already exist turbomachinery-based conversion systems that operate at these temperatures with efficiencies of >30% [2]. 1414 Degrees is a company based out of Adelaide, Australia that designed and is currently testing a silicon-based thermal energy storage system utilizing an air Brayton cycle to convert the heat back to electricity [3]. Although turbomachinery is a very mature, developed technology, a more suitable generation system for a relatively small stationary energy storage device may be a thermophotovoltaic converter.

Thermophotovoltaic systems (TPVs) are solid-state heat engines capable of converting heat to electricity through the emission and conversion of thermal radiation. These systems utilize a spectrally selective thermal emitter material heated to excess of 1000°C and capture the radiant photon emission using a photovoltaic cell. They are solid state systems that require very little maintenance and are theoretically capable of very high efficiencies. By integrating these electrical generators in the place of traditional turbomachinery, low costs and high robustness could be achieved [4]. The rest of the analysis presented in this report is for a PC-TES system using silicon as the storage medium with a TPV converter. The details of how such a system would be constructed are not included, as there is significant thermal and materials engineering research still required for a full working system.

**Analysis:**

Carbon Reductions:

As of 2016, 47% of the electricity generated by the ISO-NE grid was produced by some form of non-emitting source (9% renewables, 7% hydro, 31% nuclear) [5]. According to daily grid loading data from ISO-NE averaged over 2017 [6], minimum hourly load is 78% of the average daily load.

For the following preliminary analysis, two simplifying assumptions were made. First, that the grid will be required to use clean energy sources in favor of emitting ones (i.e. if less than 47% of the grid is in use, only non-emitting sources will be used). This can make economic sense as renewable sources typically do not have fueling costs, but new renewable installations are not always owned by major utility companies. To address this, Massachusetts has a law, 220 C.M.R. § 8.00, that requires utility providers to purchase renewable electricity at the same price that they could produce it using their existing facilities (i.e. gas turbines) [7]. This law was updated in 2009, but is a state implementation of the federal Public Utility Regulatory Policies Act of 1978. The second assumption is that under average grid load, 47% of the electricity is produced by non-emitting sources.

Based on these two assumptions and the data from ISO-NE [6], the 2017 grid operating under average load was 47% non-emitting, peak load was 40% non-emitting, and the grid operating under minimum load was 60% non-emitting. Assuming that MIT currently withdraws electricity from the grid at the average loading energy mix, drawing power from the minimum loading energy mix and storing it would result in a 42% reduction in effective $CO_2$ intensity of grid purchased electricity, from 328 kg$CO_2$/MWh$_e$ to 190 kg$CO_2$/MWh$_e$.

When looking to future carbon-emission reduction potential, two cases were considered. The first case assumes that the renewable portion of the ISO-NE grid increases by 1% per year and nuclear capacity remains constant at 31%. This case would result in a completely carbon free ISO-NE grid by 2069. By storing off peak electricity and using the analysis presented above,

MIT's electricity consumption would be carbon free by 2048.

The second case considers the more realistic assumption that nuclear capacity will decrease. This case assumes the same 1% renewables increase per year, but also accounts for a 1% per year decrease in nuclear capacity starting in 2030, resulting in 0% nuclear by 2061. This restriction results in a carbon free grid by 2101 and a carbon free electricity date of 2079 for MIT.

Costs:

Actually estimating the cost of installing such a system is challenging, but certain portions of the costs of this thermal approach to energy storage can be directly compared to the nearest alternative, lithium-ion batteries. With a 40% efficient heat-to-electric system, the volumetric electrical energy density of phase change is 513 kWh$_e$/m$^3$ for molten silicon and 448 kWh$_e$/m$^3$ for alumina with material costs of $7.5/kWh$_e$ and $3.5/kWh$_e$, respectively. These costs don't include the additional systems necessary for a standalone electricity storage system, but for general comparison lithium ion batteries have a volumetric electrical energy density of about 400 kWh$_e$/m$^3$ at a cost of around $200/kWh$_e$ [8].

To more carefully evaluate the economic viability of the proposed TPV system plus thermal storage when compared to lithium-ion batteries, a simplified levelized cost of storage (LCOS) was calculated for both devices. This metric takes into account the initial capital investment as well as the costs associated with purchasing electricity during off-peak hours for the lifetime of the system. It is important to note that the capital costs used for this comparison include only the storage device itself, the cost of inverters and additional supporting components was assumed to be similar for both the TPV and battery based systems. With these simplifications, the LCOS can be calculated as

$$LCOS = \frac{(CRF \cdot E_e \cdot N \cdot C) + \left(\frac{p}{\eta} \cdot dc \cdot E_e \cdot N \cdot 365\right)}{dc \cdot E_e \cdot N \cdot 365} \quad [1]$$

in units of $/kWh$_e$ of usable stored electricity. The remaining variables are defined in Table 1 and in the following two equations. The system capital cost ($C_{tot}$) for the TPV system can be defined as

$$C_{tot} = \frac{E_e}{\eta} \cdot C_{storage} + A_{cell} \cdot c_{cell} \quad [2]$$

To simply account for the economics of the proposed system over its 20-year lifespan, the capital recovery factor (*CRF*) was calculated for both systems using the following equation

$$CRF = \frac{r(1+r)^N}{(1+r)^N - 1} \quad [3]$$

**Table 1:** Variable descriptions and values for TPV system and Lithium-ion battery

| Variable | Description | TPV System | Lithium-Ion Battery |
|---|---|---|---|
| $CRF$ | Capital recovery factor (Eq. 3) | 0.0872 | 0.1791 |
| $E_e$ | System usable electrical storage capacity (kWh$_s$) | 1 | 1 |
| $N$ | System lifetime (years) | 20 | 7 |
| $C_{tot}$ | Total capital cost per unit storage capacity ($/kWh$_s$) | See Eqn. 2, Fig. 1a | $275 [9] |
| $p$ | Off-peak electricity price ($/kWh$_e$) | $0.029 [6] | $0.029 |
| $\eta$ | System efficiency ( - ) | See Fig. 1c | 0.85 |
| $dc$ | Discharge cycles per day (cycles/day) | 1 | 1 |
| $C_{storage}$ | Thermal storage medium+container cost ($/kWh$_t$) | $10 | - |
| $c_{cell}$ | Silicon cell cost per area ($/m$^2$) | $200 | - |
| $A_{cell}$ | Required cell area per system capacity (m$^2$/kWh$_s$) | See Fig. 1c | - |
| $r$ | Discount rate ( - ) – 6% | 0.06 | 0.06 |

Based on this LCOS analysis, Figure 1b shows how a TPV converter coupled with a silicon PC-TES medium results in a system at half the cost of an equivalent lithium-ion battery based system. The capital cost shown in Figure 1a indicates that the initial investment is almost an order of magnitude less than the costs of lithium ion batteries. The discrepancy between the LCOS difference and the capital cost difference can be accounted for by the shorter battery lifetime and lower thermal system efficiency.

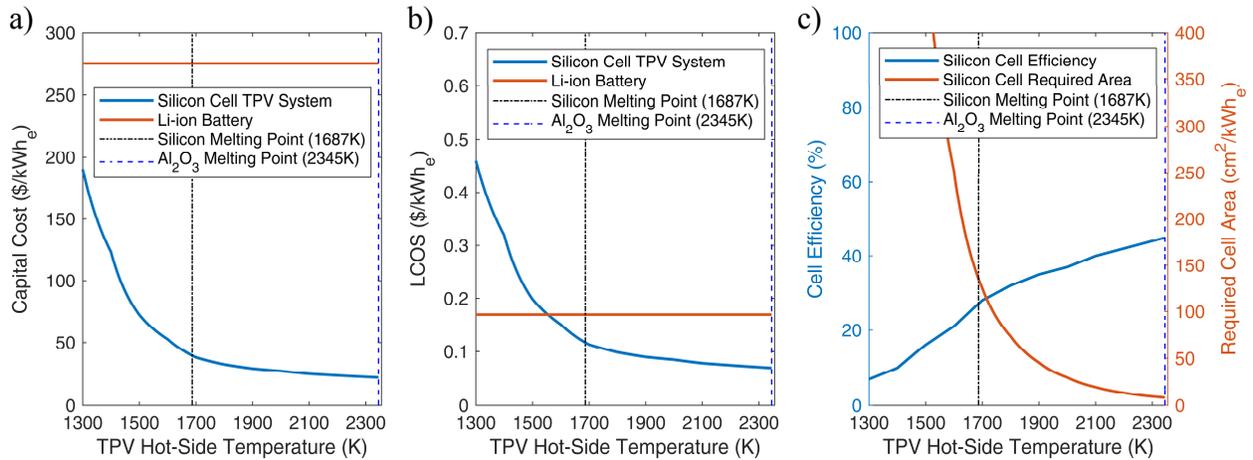

**Figure 1:** a) Capital costs comparison between TPV and lithium-ion batteries, excluding inverter and installation costs. b) Comparison of LCOS of lithium-ion batteries and TPV thermal storage at various hot-side temperatures. c) Comparison of TPV cell efficiency and required cell area at various hot-side temperatures.

Based on the capital cost requirement identified in Figure 1a, a molten silicon-based system capable of storing 2000 MWh$_e$ would cost approximately $80 million, excluding installation. If this system were implemented in 2030, the effective grid carbon intensity would be 109.8 kgCO$_2$/MWh$_e$ which could save MIT 122,000 tons of CO$_2$ emissions per year compared to 2017 emissions (assuming MIT obtains all its energy from the grid). This comes out to a cost of approximately $656 per ton of CO$_2$. Lithium ion battery based storage would cost approximately $4,100 per ton of CO$_2$, for comparison. These costs are very preliminary as they do not include additional system components (i.e. power controllers, containment facilities/buildings), but show

the clear advantages of a thermal energy storage system over battery storage.

In addition to emissions savings, there are purely financial benefits to buying electricity off peak. If the current prices for on and off-peak electricity remain constant and MIT's total energy demand moving forward is completely provided by grid electricity, storing off peak electricity ($28.88 /MWh$_e$) would save $4 million per year when compared to the average demand pricing ($34.77 /MWh$_e$)[6]. Additionally, electricity is a much more useful form of energy when compared to heat. By implementing ground source heat pumps with COPs of 4, the electricity demand of a completely grid connected MIT could be significantly less than the current total energy demand, further saving carbon and money.

Time:

Due to the experimental and cutting edge nature of this technology, several years may be required to develop this technology into a working system. Additionally, the advantages of storage become more pronounced as the grid moves towards a higher penetration of renewables. I believe it's viable to implement a storage system within the next 10 years.

Political, social, and ethical issues:

Energy storage as potential carbon-mitigation strategy is inherently relatively passive and unobtrusive. As intermittent sources become the norm, storage will become more important and may even be incentivized.

Scalability:

The proposed thermal storage technology is very scalable when compared to other energy storage methods. Due to the low cost of the thermal storage medium, increasing the storage capacity (kWh$_e$) increases the total cost of the system negligibly. Increased the output power capacity (W) is a more expensive, but the system as a whole remains low cost. Additionally, no exotic materials are required for the implementation of this system (i.e. lithium for li-ion batteries), so there are no materials limitations on the scalability of the system.

Storage will become increasingly important to a grid highly penetrated with renewables, but the financial benefits of purchasing off-peak electricity may decrease over time. As more storage is constructed, the need for storage decreases and as such it may come to a point where the daily electricity demand profile is relatively flat. In this situation, building more storage would not save money or decrease relative carbon emissions. For this to happen, however, enormous grid level storage capacity would have to be implemented by utilities.

Land-use:

If all MIT's energy were to be taken from the grid at night when demand is lower, stored in a silicon high-temperature phase change material, and converted back to electricity to power the university throughout the day, an average 2000 MWh$_e$ of electricity storage would be required. With the 40% efficient heat-to-electric conversion assumption, roughly 4000 m$^3$ of silicon would be required at a total cost of $15 million for the material. Assuming a 25% growth in energy demand over the next 50 years, 5000 m$^3$ of storage could fit in a 1-meter deep hole underneath a football field or in a 3 story building with a 600 m$^2$ footprint.

Lifecycle assessment:

The proposed system does not introduce any toxic or exotic chemicals or incur high manufacturing emissions, like traditional battery based systems. Minimal processing of the storage medium is required, as both bulk silicon and alumina are easily obtainable on an industrial scale. Although the manufacture of solar PV cells can be energy intensive, this system would use many fewer cells when compared to a PV farm. Silicon PV cells have already become commonplace in the movement to de-carbonize the US grid, and utilizing some of these cells for a storage application would not increase manufacturing emissions significantly. Additionally, this thermal storage approach does not require any of the rare elements (lithium, cobalt) currently required for battery manufacture. The emissions associated with the mining, purification, and transportation/importation of these rare elements can be eliminated with a thermal storage system.

Risks/uncertainty:

The risks associated with this proposed project can be broken down into three categories: financial risks, safety risks, carbon emissions risks.

The financial risks associated with the project primarily stem from the experimental status of the system. There will be failures and components of the system that will not work properly the first time around, and in that sense the project may end up costing more than initially planned. Additionally, if the lifetime of the proposed system is less than designed, the relative cost of the system could increase. Finally, if the off-peak and on-peak pricing of electricity changes significantly, some of the economic benefits of purchasing electricity off-peak could be lost.

Safety risks include any risks associated with the storage and electrical heating of high-temperature materials. The temperatures the proposed system would operate at would exceed 1400°C, but all hot components would be completely self-contained with little to no maintenance or human interaction. Additionally, there would be no explosion or over pressurization risk commonly associated with steam-based systems. When taking power from the grid, there is also a risk associated with the high voltages associated with mains power.

Finally, carbon emissions risks include the risk that installing this system would not reduce MIT's carbon output as much as claimed. This risk may be realized if utilities fail to meet the Renewable Energy Portfolio Standard or if an agreement is made between the ISO-NE and Hydro Quebec for grid-level storage. The proposed technology is most advantageous when significantly grid-level storage is not available and MIT can take advantage of the daily fluctuations in electricity demand.

**Conclusion:**

Energy storage and arbitrage is not the most traditional way to reduce greenhouse emissions, but it is a viable option for MIT to achieve its carbon neutrality goals. Without abundant renewable energy resources (i.e. land/roof area), urban campuses like MIT are beholden to the carbon intensity of local grids. Although Massachusetts does have legislation mandating the increased penetration of renewables, the current progress is far too slow for MIT to reach its goals through grid decarbonization alone. Energy storage would allow MIT to take advantage of the existing non-emitting resources, and effectively achieve carbon neutrality far before the grid is carbon-free. The proposed high-temperature thermal energy storage system may provide a viable, lower cost solution to the typically high cost of battery based storage. Further research and development still needs to be done to make this system a reality, but its potential is significant both for MIT and the energy industry as a whole. By implementing the proposed thermal energy storage system by 2030 at a cost of $80 million (excluding construction and supporting infrastructure), MIT could be carbon-free by as early as 2048.